\documentclass[11pt,a4paper]{article}
\pdfoutput=1 
\usepackage{graphicx}
\usepackage{color}
\usepackage{lineno}
\usepackage{textgreek} 
\definecolor{red}{rgb}{1,0,0}
\newcommand{\rod}[1]{\textcolor{red}{\textbf{#1}}}
\definecolor{blue}{rgb}{0,0,1}

\newif\ifdraftlayout
\draftlayoutfalse

\newif\ifjournallayout
\journallayoutfalse

\newif\ifexpandfigures
\expandfigurestrue

\ifjournallayout
  \usepackage{jinstpub} 
\else
\fi

\title{Accelerator Development at the FREIA Laboratory}

\ifjournallayout

\arxivnumber{to be inserted}

\author{R.~Ruber,}
\author[1]{A.K.~Bhattacharyya%
\note{Present affiliation European Spallation Source, Lund, Sweden.},}
\author{D.~Dancila,}
\author{T.~Ekel{\"o}f,}
\author{J.~Eriksson,}
\author{K.~Fransson,}
\author{K.~Gajewski,}
\author{V.~Goryashko,}
\author{L.~Hermansson,}
\author{M.~Jacewicz,}
\author{M.~Jobs,}
\author{{\AA}.~J{\"o}nsson,}
\author{H.~Li,}
\author{T.~Lofnes,}
\author{A.~Miyazaki,}
\author{M.~Olveg{\aa}rd,}
\author{E.~Pehlivan,}
\author{T.~Peterson,}
\author{K.~Pepitone,}
\author{A.~Rydberg,}
\author{R.~Santiago~Kern,}
\author{R.~Wedberg,}
\author{A.~Wiren,}
\author[1]{R.~Yogi,}
\author{and V.~Ziemann}

\affiliation{FREIA Laboratory, Department of Physics and Astronomy,\\
Uppsala University, Uppsala, Sweden}

\emailAdd{ruber@physics.uu.se}

\keywords{Accelerator Subsystems and Technologies,
Acceleration cavities and superconducting magnets
}

\abstract{%
The FREIA Laboratory at Uppsala University focuses on superconducting technology
and accelerator development. It actively supports the development of the 
European Spallation Source, CERN, and MAX IV, among others.
FREIA has developed test facilities for superconducting accelerator technology
such as a double-cavity horizontal test cryostat, a vertical cryostat with a 
novel magnetic field compensation scheme, and a test stand for short 
cryomodules. Accelerating cavities have been tested in the horizontal cryostat,
crab-cavities cavities in the vertical cryostat, and cryomodules for ESS on the
cryomodule test stand.
High power radio-frequency amplifier prototypes based on vacuum tube technology
were developed for driving spoke cavities. Solid-state amplifiers and
power combiners are under development for future projects.
We present the status of the FREIA Laboratory complemented with results of 
recent projects and future prospects.
}

\else

\author{
R. Ruber,
A.K.~Bhattacharyya\thanks{Present affiliation European Spallation Source, Lund, Sweden},
D. Dancila,
T. Ekel{\"o}f,
J. Eriksson,
K. Fransson, \\
K. Gajewski,
V. Goryashko,
L. Hermansson,
M. Jacewicz,
M. Jobs,
{\AA}. J{\"o}nsson, \\
H. Li,
T. Lofnes,
A. Miyazaki,
M. Olveg{\aa}rd,
E. Pehlivan, \\
T. Peterson,
K. Pepitone,
A. Rydberg,
R. Santiago Kern,
R. Wedberg, \\
A. Wiren,
R. Yogi\footnotemark[1],
V. Ziemann
\\
\\
FREIA Laboratory, \\
Department of Physics and Astronomy, \\
Uppsala University, Uppsala, Sweden
}

\fi

\ifdraftlayout
  \date{\rod{Draft Version: \today}}
\else
	\date{\today}
\fi

\begin{document}

\ifdraftlayout
  \linenumbers
  \modulolinenumbers
\fi

\maketitle

%
%
\ifjournallayout
  \flushbottom
\else

\ifexpandfigures
  \begin{center}
    \includegraphics[clip,keepaspectratio,width=0.5\textwidth]{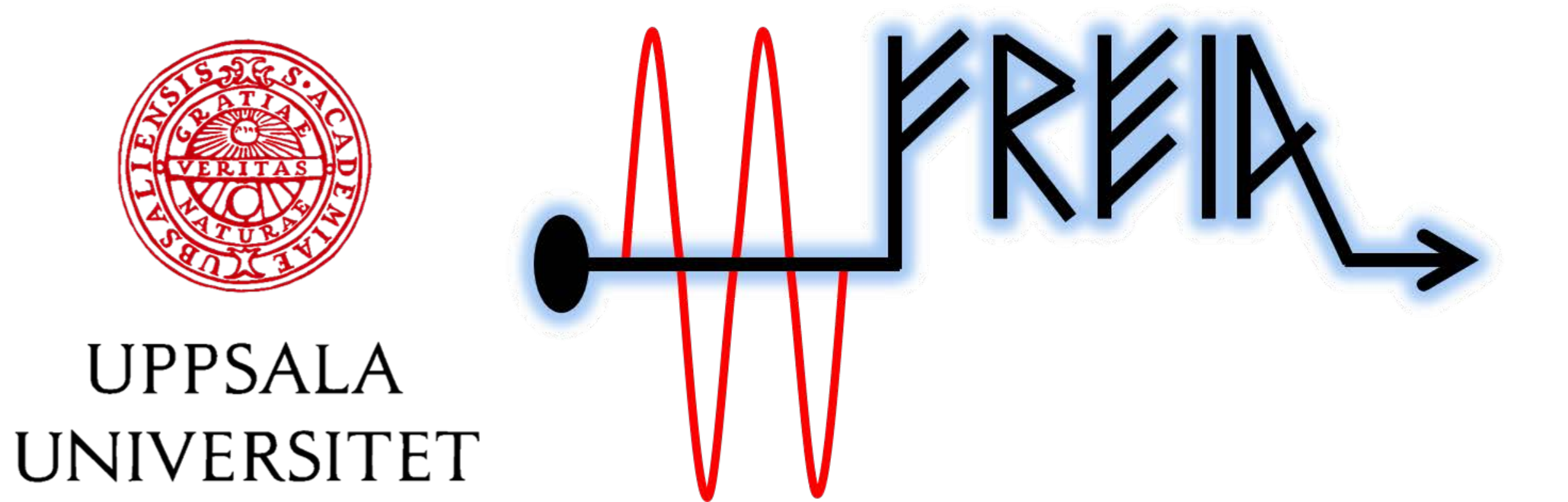}
	\end{center}
	\vspace{2\baselineskip}
\fi

\begin{abstract}

The FREIA Laboratory at Uppsala University focuses on superconducting technology
and accelerator development. It actively supports the development of the 
European Spallation Source, CERN, and MAX IV, among others.
FREIA has developed test facilities for superconducting accelerator technology
such as a double-cavity horizontal test cryostat, a vertical cryostat with a 
novel magnetic field compensation scheme, and a test stand for short 
cryomodules. Accelerating cavities have been tested in the horizontal cryostat,
crab-cavities in the vertical cryostat, and cryomodules for ESS on the
cryomodule test stand.
High power radio-frequency amplifier prototypes based on vacuum tube technology
were developed for driving spoke cavities. Solid-state amplifiers and
power combiners are under development for future projects.
We present the status of the FREIA Laboratory complemented with results of 
recent projects and future prospects.

\end{abstract}
\newpage

\fi
%
%
\section{Introduction}

In May 2009 the European research ministers agreed to build the next large
European research infrastructure, the European Spallation Source (ESS), in
Sweden. Since, at that time, the ESS had limited personnel, Swedish universities
were called upon to participate. Uppsala University was particularly suitable
to respond to this call, because it had  operated the Gustav-Werner cyclotron
since the late 1940s~\cite{GWCYC} and later the CELSIUS cooler 
ring~\cite{CELSIUS} with the WASA detector~\cite{WASA}.
In recent times Uppsala University had participated in CTF3 at CERN~\cite{TBTS}
and in FLASH~\cite{ORS} and XFEL~\cite{XFEL}, both at DESY. Moreover, experience
with cryogenics and superconducting magnets~\cite{WSCS,ATLAS} was available, 
which was particularly useful considering that ESS uses superconducting acceleration structures.

\begin{figure}[b]
\ifexpandfigures
   \centering
   \includegraphics[clip,keepaspectratio,width=\textwidth]{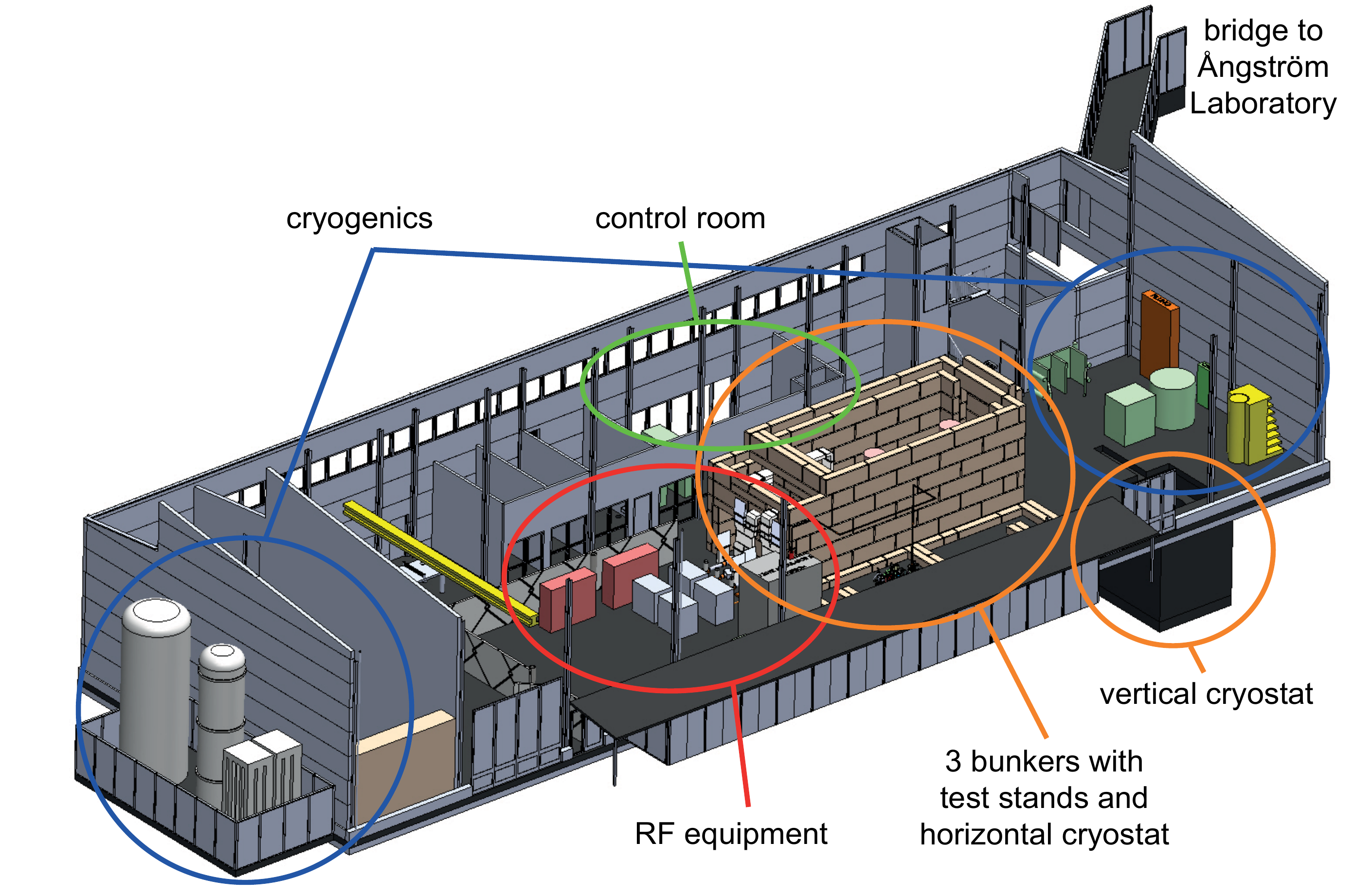}
\fi
  \caption[FREIA Laboratory]{\label{fig:freia}
  Layout of the FREIA Laboratory.
	}
\end{figure}

The scope of the collaboration between ESS and Uppsala University was agreed to
comprise the design of a radio-frequency (RF) power generation and 
distribution system, as well as testing all components, including 
superconducting RF accelerator cavities. This stimulated the construction of a
complete test-stand in Uppsala, similar to CHECHIA~\cite{CHECHIA}, 
CRYHOLAB~\cite{CRYHOLAB}, or  HoBiCat~\cite{HOBICAT}.
The focus of the new facility became spoke type superconducting RF 
cavities for which no other high-power test stand is available in Europe.
These types of superconducting cavities, also known as spoke resonator, were 
first proposed in the late 1990s~\cite{delayen1998} but the ESS will be the 
first large infrastructure to use them.
For ESS, these cavities have two spokes, resulting in three acceleration gaps,
and are referred to as double-spoke cavities. They accelerate the proton beam 
from 90\,MeV kinetic energy at the output of the normal conducting section of
the accelerator to 220\,MeV at the input of the elliptical cavity section of 
the accelerator.

The new facility was formally established in 2011 and named FREIA (Facility for
Research Instrumentation and Accelerators). The layout of the laboratory is 
shown in figure~\ref{fig:freia} with bunkers for the horizontal cryostat in the
center of a newly erected 1000\,m$^2$ building. The site of the laboratory was
decided to be placed adjacent to the {\AA}ngstr{\"o}m Laboratory, the present 
site of the physics, engineering and chemistry departments. Construction of the
new laboratory building was completed in 2013.

The basic layout of the test areas includes the superconducting accelerating
cavities, high-power RF source, RF distribution, and controls system. The 
central part of the laboratory comprises a horizontal cryostat that can
accommodate one or two superconducting cavities. The cryostat has to be located
in a concrete bunker for radiation protection purposes and is cooled by a
cryogenic system providing liquid helium and liquid nitrogen; all is
supervised by an integrated controls and monitoring system. Additional space is
allocated inside the bunker for the testing of cryomodules, which are 
specialized cryostat-modules, with one or multiple accelerating cavities, that 
are installed in the final accelerator itself.
The RF power stations are placed outside of the bunker, adjacent to its western
side. Due to an architectural intervention, the cryogenic system ended up on
the eastern end of the building while the helium compressor and gas storage 
ended up on the western end. Control room, workshops, and storage areas are on
the ground level with office space above.

An optional vertical cryostat was considered from the beginning as the simplest
way for future testing of superconducting cavities. This idea was later updated
for multi-purpose superconducting cavity and magnet testing, and the cryostat
was finally added in 2019.

In the following sections we walk the reader through all parts, starting with
the cryostats, which occupy a central role for testing components, and the 
associated cryogenics system. This is followed by a section on the power
generation and distribution, which served as a prototype for part of the ESS
system. Then we address the test infrastructure and present recent results.
Finally, we discuss upcoming and future activities.

%
%
\section{Cryostats}

The horizontal cryostat ``Hnoss\footnote{In Nordic mythology, Hnoss and Gersemi
are daughters of the goddess Freia.}'' is dedicated to test superconducting
cavities and located inside a radiation protection bunker next to a test stand 
for testing cryomodules for the ESS. 
Our vertical cryostat ``Gersemi\footnotemark[1],'' is located in a 
vertical shaft and used to test both superconducting cavities and magnets.

\subsection{``Hnoss'', The Horizontal Cryostat}

\begin{figure}[b]
\ifexpandfigures
   \centering
   \includegraphics[clip,keepaspectratio,width=0.5\textwidth]{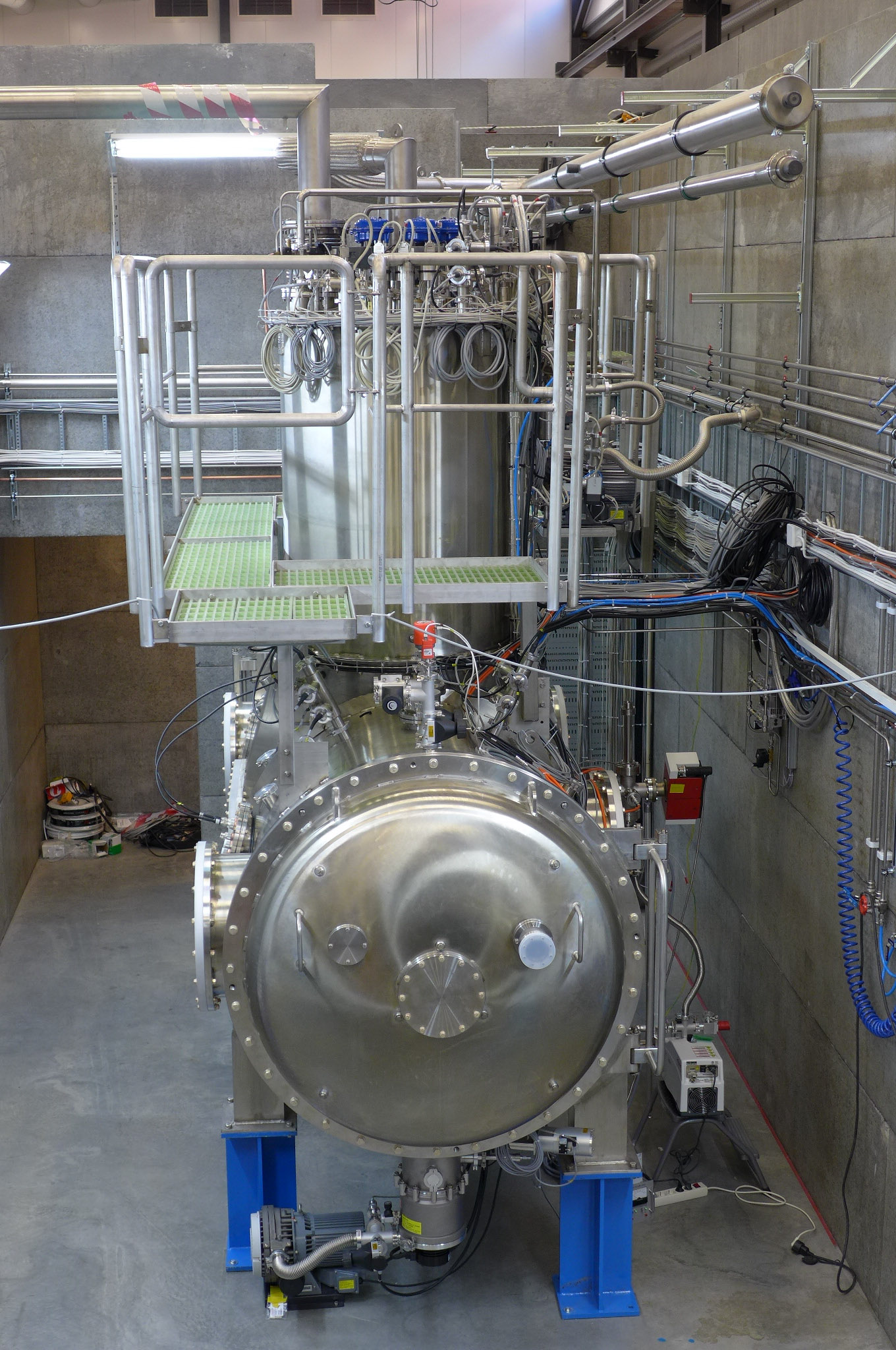}
\fi
  \caption[HNOSS]{\label{fig:hnoss}
  The Hnoss horizontal cryostat, looking towards one of its loading doors, with
	the cryogenic valve box, surrounded by a maintenance platform, on its top.
  Insulation vacuum pumps can be seen below.
	}
\end{figure}

\begin{figure}[b]
\ifexpandfigures
   \centering
   \includegraphics[clip,keepaspectratio,width=\textwidth]{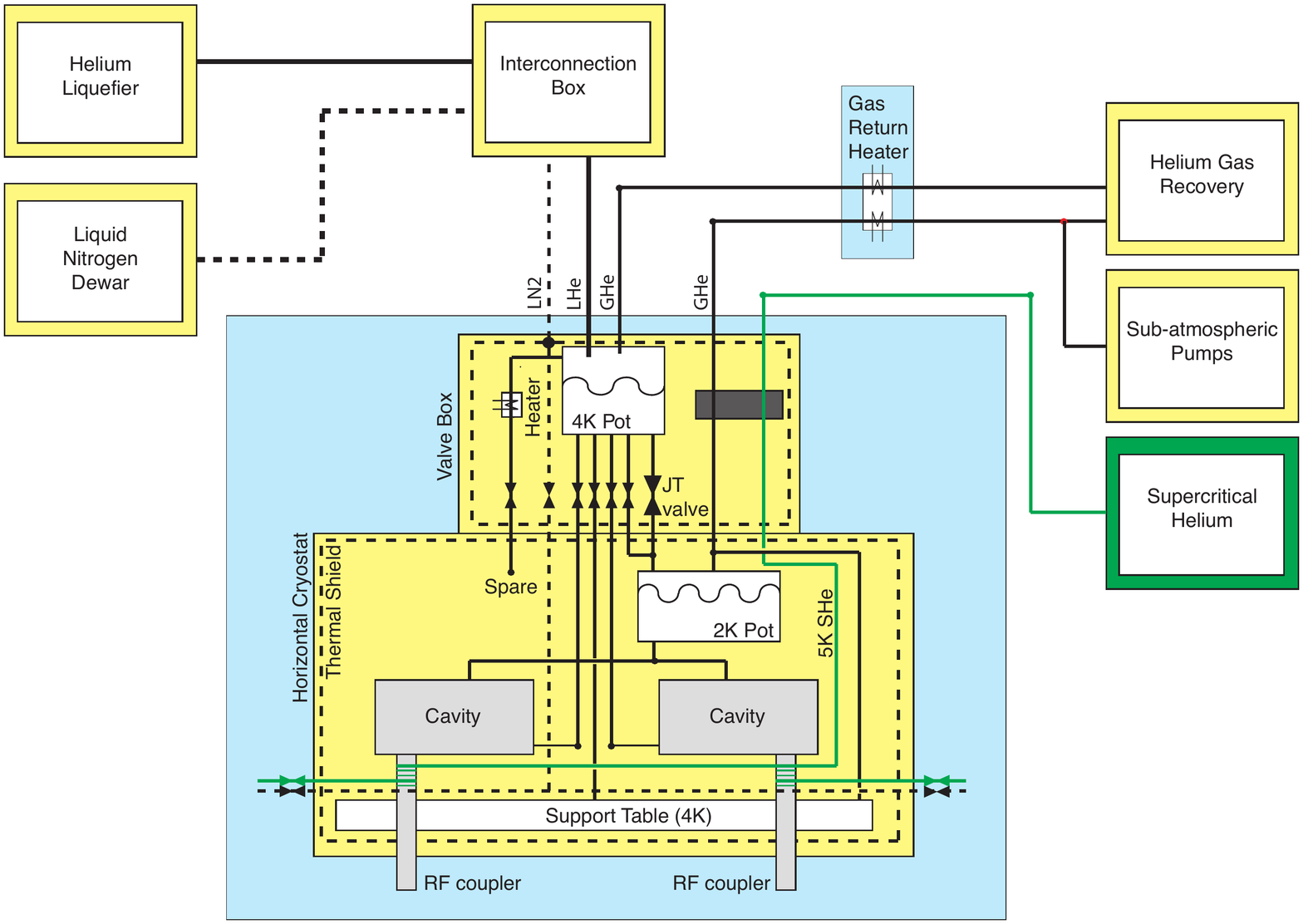}
\fi
  \caption[HNOSS Cryogenics Scheme]{\label{fig:hnoss-circuit}
  The cryogenic scheme of Hnoss with the valve box including the 4K pot on top
	of the horizontal cryostat with the cavities. Liquid nitrogen (LN2, dashed) 
	is used to cool the thermal shield and RF couplers and liquid helium 
	(LHe, solid) to cool the cavities and the support table.
	Gas helium return lines are marked ``GHe''.
	The supercritical helium line (green) as alternative cooling for the RF
	couplers is marked ``5K SHe''.
	}
\end{figure}

The outer housing of Hnoss is a stainless steel cylinder with large doors at
its ends and a valve box on top, as shown in figure~\ref{fig:hnoss}. Multiple
vacuum flanges are available to mount feedthroughs for signal cables and to
connect power couplers for the RF, either from the left, right, top or bottom.
With its inner length of 3.3\,m and diameter of 1.2\,m, Hnoss is large enough 
to accept two superconducting cavities with their individual liquid helium
reservoirs and with power couplers installed. In particular, Hnoss can 
accommodate two ESS-type 352\,MHz double-spoke cavities, but also two 704\,MHz
ESS-type elliptical cavities or two TESLA/ILC-type 1.3\,GHz 
cavities~\cite{junquera2013}.

Inside the vessel a mu-metal sheet reduces the Earth-magnetic field
by a factor of~5 and an aluminium sheet functions as thermal-radiation shield
while cooled with liquid nitrogen to about 80\,K~\cite{santiago2014}. The
superconducting cavities, or any other devices under test, are cooled by liquid
helium to temperatures between 1.8 and 4.2\,K. The cavities are either suspended
from the outer vessel wall by tie-rods or supported by a liquid helium cooled
table that is placed at the bottom of the cryostat. This table is equipped with
rails to simplify installation work, but it can also be removed to create more
space for larger cavities.

An interconnection box inside the bunker routes the liquid helium, arriving
from the liquefier at about 1.2\,bar, to the valve box on top of the Hnoss
cryostat. This cooling circuit is shown in figure~\ref{fig:hnoss-circuit}.
The yellow boxes represent different sub-systems. Liquid helium and nitrogen
arrive through the interconnection box into the valve box which sits on top
of the horizontal cryostat itself.
Inside the valve box the helium enters a small storage dewar, called
the 4K pot, from where it can be routed directly to the cavities (or any other
device under test), during initial cool down at 4\,K, and to the support table.
A third route passes the 4\,K Helium via a heat exchanger to a Joule-Thomson
valve, where the helium temperature is lowered by isenthalpic expansion.
The resulting helium, at temperatures down to 1.8\,K, is collected in 
a second dewar, called the 2K pot, from where it is delivered to the 
cavities~\cite{chevalier2014}. A bypass line for the Joule-Thompson valve 
allows us to operate fully in 4\,K mode. After having cooled the cavities, the
returning gas passes the cold side of the previously mentioned heat-exchanger
to pre-cool helium on its way to the Joule-Thomson valve. The return gas leaves
the valve box through a port on the top from where it passes a heater system,
located outside the bunker, which increases the gas temperature to ambient
temperature before entering sub-atmospheric pumps that lower the pressure to
values required for the desired outlet temperature at the Joule-Thomson valve.

The valve box also contains circuits for liquid nitrogen to cool the thermal
shield of the cryostat. The RF couplers of the cavities can be cooled either by
the liquid nitrogen or by 5\,K supercritical helium.
The supercritical helium is produced in a secondary helium circuit that is 
cooled by an additional heat exchanger on the exhaust of the 2K pot. The helium
gas in this secondary circuit is fully separated from the helium gas circuit of
the cryogenic plant.

\subsection{Cryomodule Test Stand}

\begin{figure}[b]
\ifexpandfigures
   \centering
   \includegraphics[clip,keepaspectratio,width=\textwidth]{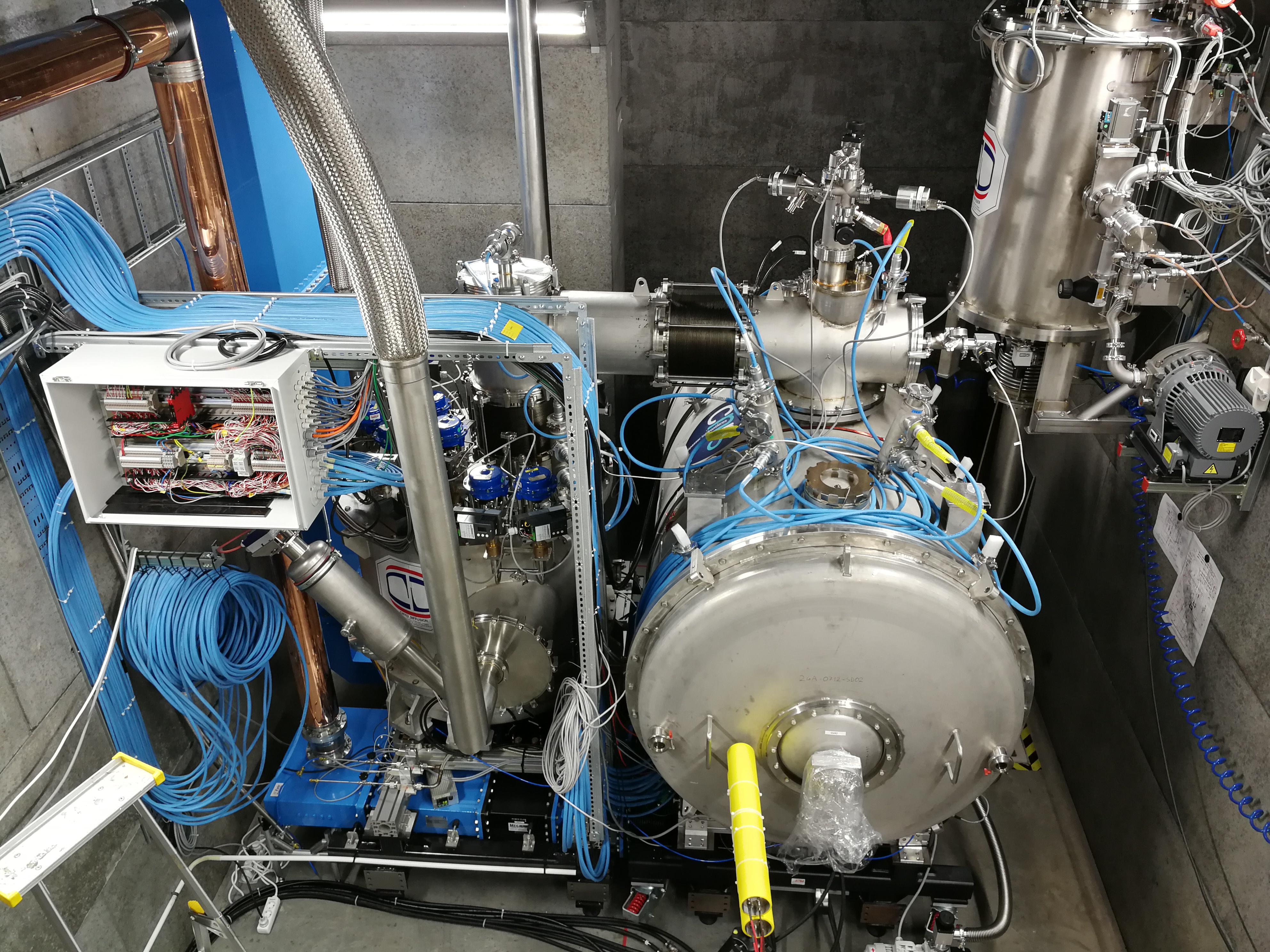}
\fi
  \caption[Cryomodule]{\label{fig:cryomodule}
  The test stand with an ESS cryomodule shown at the right and the valve box at
	the left. Liquid helium and liquid nitrogen arrive through the cryogenic
	interconnection box visible at the top-right.
	RF power arrives through the coaxial waveguide (copper coloured) and
	rectangular waveguide (blue) shown at the far-left.
	}
\end{figure}

Inside the bunker, adjacent to Hnoss, space is available to test ESS 
cryomodules. The same interconnection box that also supplies Hnoss, delivers
liquid helium and liquid nitrogen to a valve box that is located next to the
cryomodule; a setup that mimics the installation in the ESS accelerator. 
Figure~\ref{fig:cryomodule} shows a prototype valve box and cryomodule for 
spoke cavities installed in the bunker. A similar cryogenic flow principle is
used as for Hnoss: the 2\,K helium from the valve box is routed to the 
cryomodule, where it cools two spoke-cavities. After leaving the cryomodule, 
the return gas passes through the same heater and the same sub-atmospheric 
pumps that are used for Hnoss in order to lower the pressure to values needed 
to operate the Joule-Thomson valve in the valve box.

Valve box and cryomodule are rolled into position using a trolley by entering
the bunker through a demountable wall at the backside of the bunker, opposite 
to the side facing the RF stations. For the series acceptance testing of ESS
cryomodules the same valve box will remain in place for all tests.

\subsection{``Gersemi'', The Vertical Cryostat}

\begin{figure}[b]
\ifexpandfigures
   \centering
   \includegraphics[clip,keepaspectratio,width=0.5\textwidth]{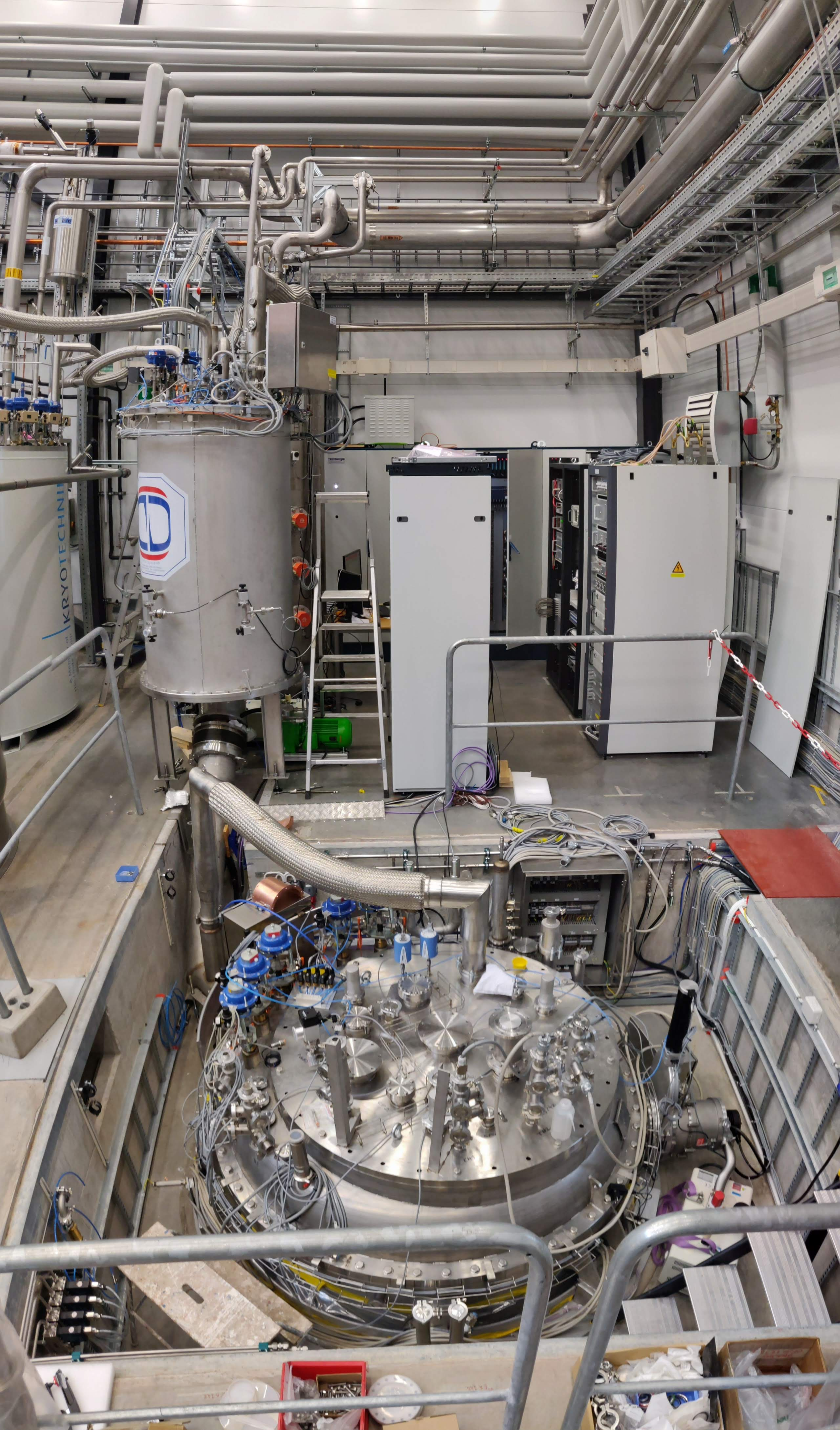}
\fi
  \caption[Gersemi]{\label{fig:gersemi}
  The Gersemi vertical cryostat inside its shaft with the valve box at the rear
	left. A transfer line for liquid helium and  nitrogen connects from the 
	bottom of the valve box to the left side of the cryostat and is partly hidden
	by the flexible transfer line for the return gas.
  The racks with power supplies and controls equipment are at the rear right.	
	}
\end{figure}

\begin{figure}[b]
\ifexpandfigures
   \centering
   \includegraphics[clip,keepaspectratio,width=\textwidth]{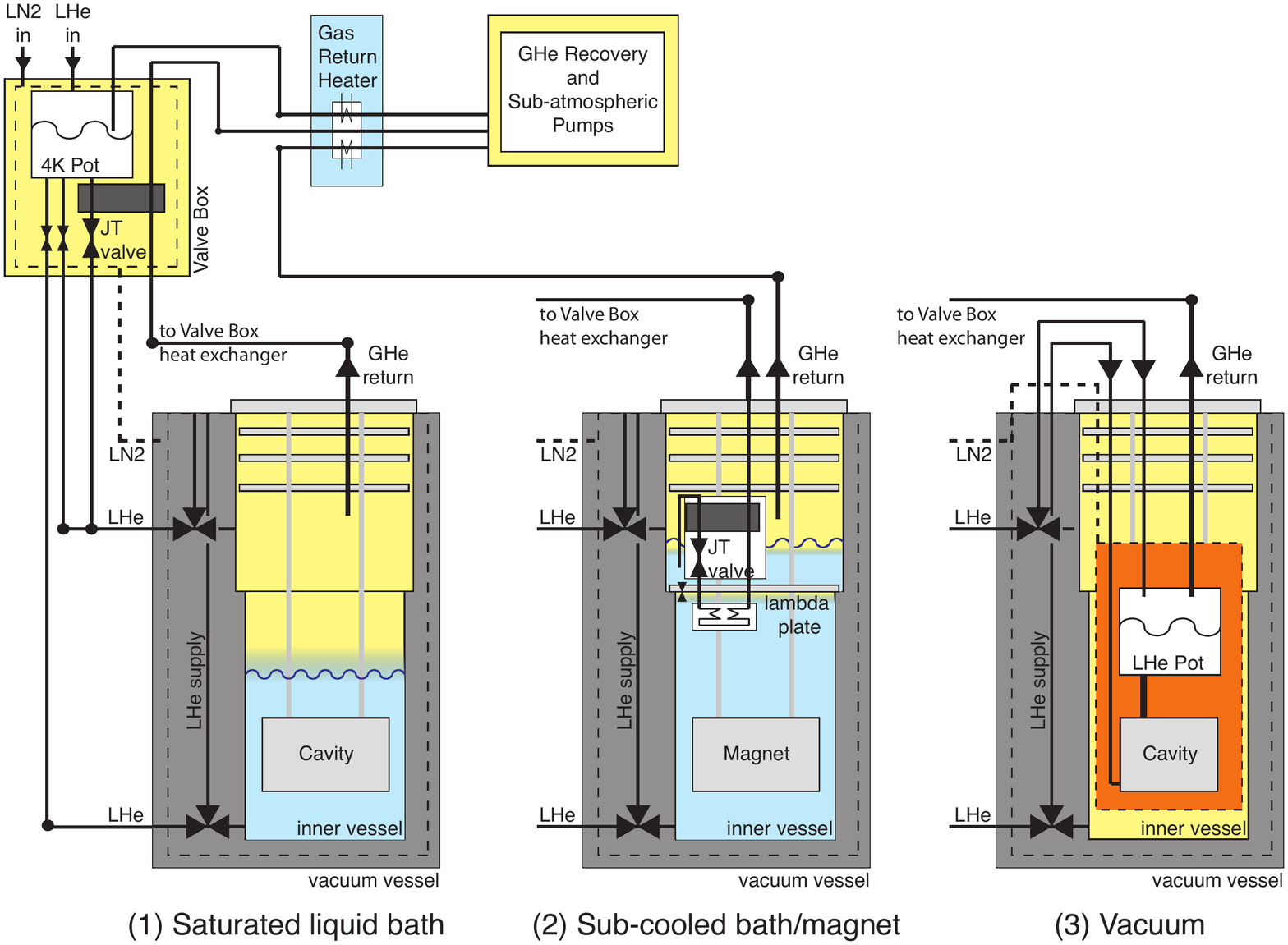}
\fi
  \caption[Gersemi Inserts]{\label{fig:gersemi-inserts}
  Gersemi's three operational modes and insert types: liquid, magnet, and 
	vacuum.
	Liquid helium and nitrogen from the valve box (top left) enter 
	the cryostat (gray) from the left, where valves are used to orchestrate the
	helium flow for the three operation modes.
	The magnet mode uses two gas helium return lines: one for the 
	helium bath above and one for below the lambda plate.
	}
\end{figure}

The construction of Gersemi resembles that of Hnoss and is connected to the 
same sub-atmospheric pumps. Only, in Gersemi, the valve box with the 
Joule-Thomson valve to provide 2\,K helium shown on the top-left in 
figure~\ref{fig:gersemi} is standing next to a 5\,m deep shaft that has a 
diameter of 2\,m. This shaft contains the stainless steel vessel of the vertical
cryostat itself, in which to test either superconducting cavities or magnets.
The top flange of the cryostat is visible near the bottom of 
figure~\ref{fig:gersemi}.
The space inside the cryostat is limited to a height of 3.05\,m and a diameter 
of 1.1\,m. Moreover, in order to test magnets in a pressurized 2\,K helium bath,
a so-called lambda-plate separate the 2\,K helium bath from a 4\,K bath above. 
The available height below the lambda-plate is 2.65\,m.

Gersemi supports three operational modes: (1) saturated liquid bath, (2) 
sub-cooled bath for magnets, and (3) vacuum. 
So-called ``inserts'' are used for testing a device with each operational mode
having its own insert type shown in figure~\ref{fig:gersemi-inserts}. An insert
is in fact the top flange of the cryostat with a mechanical support for 
suspending the device to be tested. The insert includes feedthroughs for all
necessary connections: cryogenics, vacuum, signals etc. 

In mode 1, bare cavities without surrounding helium container are directly 
placed in a saturated liquid bath with temperatures in the range of 1.8 to 
4.5\,K, depending on the pressure.
The liquid helium supply is organized such that the cryostat is cooled down
and filled from the bottom, and then shifts to the top during steady state
operation. The cold helium gas returns through a heat exchanger in the valve 
box and connects either to the 2\,K sub-atmospheric pumping system or via a 
bypass to the gas recovery system.

Mode 2 is used for superconducting magnets.
The magnet is suspended below the lambda-plate in a pressurized 2\,K helium 
bath, while above the lambda-plate there is a saturated 4\,K helium bath. 
As in mode 1, the liquid helium supply is organized such that the cryostat is
cooled down and filled through the bottom, and then shifts to the top during
steady state operation. 
To cool the bath below the lambda-plate, a fraction of the 4\,K helium above 
the lambda-plate is pumped by the sub-atmospheric pumps through a heat exchanger
and a Joule-Thomson valve. The helium continues through a second heat exchanger
cooling the helium bath below the lambda-plate before flowing back through the
first heat exchanger cooling the incoming 4\,K helium~\cite{thermeau2019}.
A separate cold gas return line, passing through the valve box, is used for the
4\,K bath itself.

Mode 3 is similar to the operation in Hnoss, where superconducting cavities, or
any other devices, are equipped with a liquid helium vessel. 
In this mode of operation, the surrounding test volume of the cryostat is
evacuated, hence the naming vacuum mode.
The liquid helium supply is organized through valves in the cryostat that
redirect the liquid helium through a short transfer line to the top of the 
insert and from there to the helium reservoir of the device being tested. 
The cold helium gas return connects through a heat exchanger in the valve box
either to the 2\,K sub-atmospheric pumping system or to a bypass to the gas
recovery system.
An additional transfer line can be used to supply liquid nitrogen if a separate
thermal screen is installed for the insert.
As this mode of operation is equivalent to testing in Hnoss, the hardware of
the insert and transfer lines has not yet been constructed.

Initially, a removable magnetic shield was foreseen outside the vacuum vessel.
It was planned to use the shield when testing superconducting cavities and to
remove it when testing superconducting magnets. Since such a magnetic shield 
would be rather fragile, we replaced it by an active earth-field compensation
magnetic system consisting of three horizontal and three vertical coils that
surround the vessel of the cryostat~\cite{freia:report1801}. Three horizontal
coils, spaced 1\,m apart, produce a reasonably constant vertical field 
component, provided they are excited in a 1, 2/3, 1 pattern. We found this
configuration by optimizing the field flatness over 1.6\,m and picked an integer
wiring ratio with small numbers of 8 and 12 windings for the center and 
end-coils, respectively. The three vertically mounted coils have a height of 
3.7\,m and can be powered independently to provide any horizontal field 
component. The field quality within a circle of 0.4\,m radius is below 
4\,{\textmu}T at an excitation of 60\,{\textmu}T, or below 
3\,\%~\cite{ziemann2020}.

%
%
\section{Cryogenic System}

\begin{figure}[b]
\ifexpandfigures
   \centering
   \includegraphics[clip,keepaspectratio,width=\textwidth]{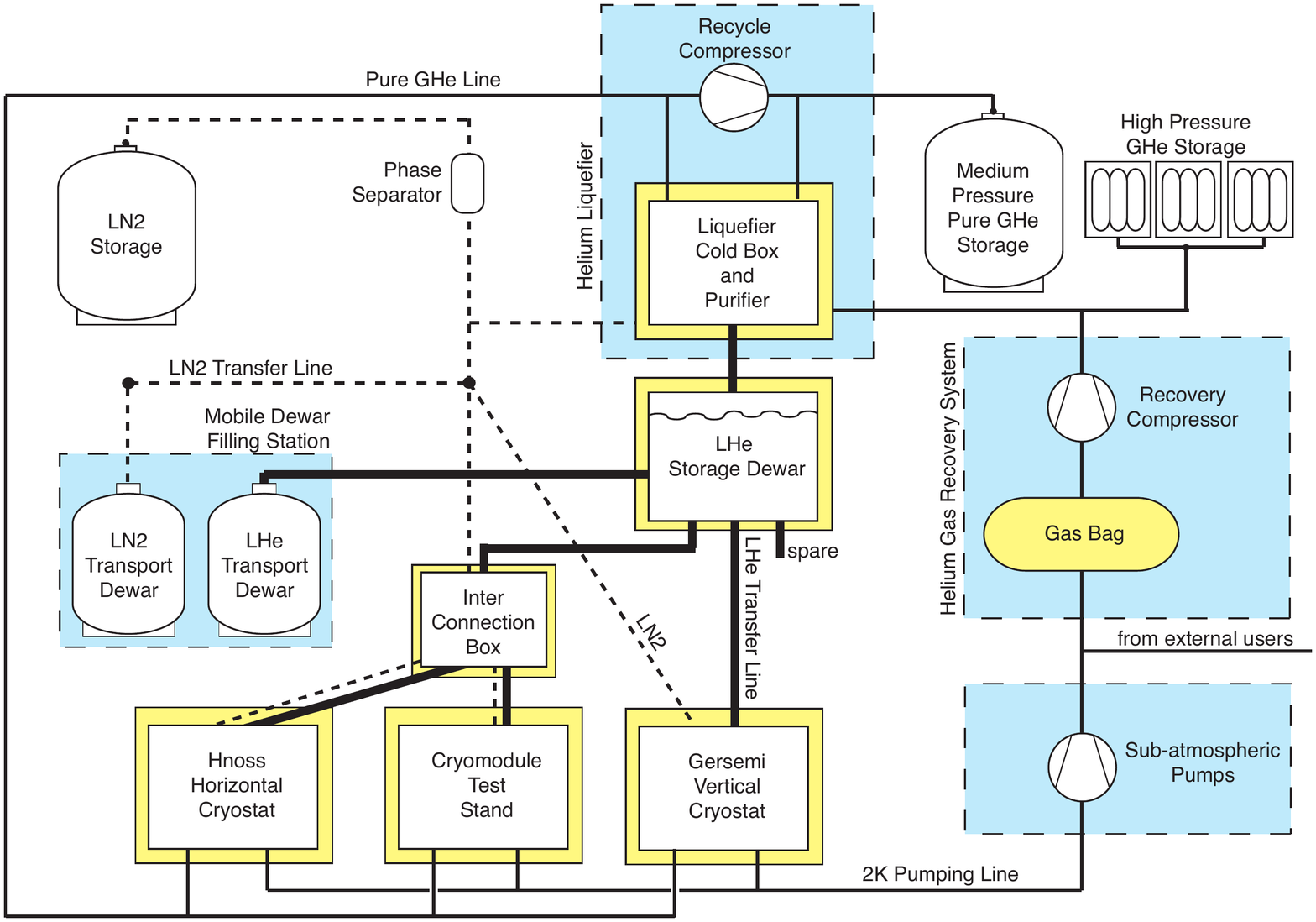}
\fi
  \caption[Cryogenic System]{\label{fig:cryosystem}
	Schematic view of the cryogenic system showing the helium liquefier (top
	centre) delivering liquid helium via the storage dewar either to the
	interconnection box, and to Hnoss (figure~\ref{fig:hnoss} and
	\ref{fig:hnoss-circuit}) and the cryomodule (figure~\ref{fig:cryomodule}),
	or to Gersemi (figure~\ref{fig:gersemi} and \ref{fig:gersemi-inserts}).
	The sub-atmospheric pumps (bottom right) facilitate operation at 2\,K.
	Liquid helium and gas helium connections are shown as solid lines
	and liquid nitrogen as dashed lines.
	}
\end{figure}

A schematic overview of the liquid nitrogen and liquid helium cryogenic system
is shown in figure~\ref{fig:cryosystem}. A helium liquefier supplies an
intermediate storage dewar from which the liquid helium is distributed to
Gersemi and to an interconnection box which in turn supplies Hnoss and the
cryomodule test stand \cite{freia:report1503}. The helium gas returns directly
to the recycle compressor of the helium liquefier, or to the 
sub-atmospheric pumping and gas recovery systems.
In parallel, liquid nitrogen is used for pre-cooling of the liquefier cold box
and cryostat thermal shields. 
The main parameters of the cryogenic system are listed in 
table~\ref{tb:cryosystem}.

Over 50,000\,l liquid nitrogen and 8,000\,l liquid helium are distributed annually
to external users at the university, other government agencies and private companies nearby.
The liquid nitrogen is stored in a 20,000\,l medium pressure dewar at some 60\,m
distance from the helium liquefier, therefore a phase-separator is used to 
reduce the pressure and remove any gas nitrogen at the end of the 
transfer-line.

\begin{table}[t]
\centering
\caption[Cryogenic Parameters]{Main parameters of the cryogenic system.}
\label{tb:cryosystem}
\begin{tabular}{l l}
\multicolumn{2}{l}{ } \\
Liquefier                        & \\
- nominal capacity               & 140\,l/h at 1.15\,bar \\
                                 & 150\,l/h at 1.23\,bar \\
- pre-cooling                    & 70\,l/h liquid nitrogen \\
- purifier minimum inlet pressure   & 25\,bar \\
Liquid helium storage dewar      & 2000\,l \\
Recovery gas balloon             & 100\,m$^3$ \\
Recovery compressors             & $>$75\,m$^3$/h at 200\,bar \\
High pressure storage            & 14.4\,m$^3$ at 200\,bar \\
Sub-atmospheric pumps            & 3.2\,g/s at 10\,mbar \\
                                 & 4.3\,g/s at 15\,mbar \\
\end{tabular}
\end{table}

\subsection{Helium Liquefaction}

The helium liquefier system consists of a recycle compressor and a cold box.
The recycle compressor is a Kaeser screw compressor with a maximum discharge
pressure of 14\,bar. The cold box is a Linde L140 type containing the heat
exchangers and expanders that cool down and liquefy the helium gas.
A fraction of the gas flow from the recycle compressor is led through two
turbines where the gas is cooled by adiabatic expansion. This gas flow is then
used to cool down the heat exchangers which in turn cool down the remaining
fraction of the gas flow. This high pressure cold gas then passes through a 
Joule-Thompson valve in an isenthalpic expansion process and is liquefied and
stored in a 2000\,l dewar.
A liquefaction rate exceeding 150\,l/h was reached with liquid nitrogen 
pre-cooling of the first heat exchanger in the cold box, and about half that
rate without liquid nitrogen pre-cooling.

To avoid clogging of the cold box by frozen impurities during long-term 
operation, it is equipped with adsorbers that freeze out impurities at 80\,K 
and 20\,K. To prevent warming-up the whole cold box to regenerate the 80\,K
adsorber, it is equipped with valves for in-line regeneration.

The 2000\,l storage dewar acts as a buffer in case a different liquid helium
throughput rate is required than provided by the liquefier cold box. 
The control software can adjust the liquefaction rate to maintain a constant
liquid level in the storage dewar by adjusting the gas pressure at the inlet of
the cold box. The storage dewar has four connections for the distribution of
liquid helium. One is used to fill mobile dewars and the other three have 
integrated cryogenic valves, one of which links to the interconnection box 
(and from there to Hnoss and the cryomodule test stand), another links to
Gersemi, while the last is a spare.

The helium gas returning from the cryostats and external users is fed back to
the recycle compressor. A medium pressure tank serves as buffer for the pure 
gas helium in case the incoming flow to the recycle compressor is higher than
required by the cold box. If more gas is needed, it can be provided from a high
pressure gas storage. Before being used, this gas it is sent through a 
freeze-out purifier included in the liquefier cold box to remove any
impurities.

\subsection{Sub-atmospheric Pumping}

The sub-atmospheric pumping system is used to lower the pressure of the liquid
helium bath in the cryostats and thereby the temperature. The dominant 
constraint for the minimum temperature is the pressure that can be reached by
the pumps, which in turn depends on the return mass flow at the lowest pressure.
The system has a flow capacity of 10\,g/s at 200\,mbar down to 3.2\,g/s at 
10\,mbar (room temperature). Assuming that the flow capacity is roughly 
proportional to the pressure, the system supports a flow of 4\,g/s at 31\,mbar 
which corresponds to about 100\,W cooling power at 2.0\,K.

The pumping system is an integrated set of roots and dry-pumps.
The first stage consists of two roots pumps in parallel, the second stage 
consists of a single roots pump, and the third stage consists of two dry 
roughing pump sets connected in parallel. Water cooled heat exchangers
refrigerate the gas between the stages. Pressure regulation is achieved either
by regulating the speed of the pumps or controlled opening of a butterfly 
valve between pumps and liquid helium bath.

The helium gas from the sub-atmospheric pumping system is recovered in a 
100\,m$^3$ balloon at atmospheric pressure. The same gas balloon is also fed 
with helium gas recovered from colleagues using mobile dewars in adjacent
faculty buildings up to 1\,km distance (marked as ``from external users'' in
figure~\ref{fig:cryosystem}). The gas balloon is emptied by a set of recovery
compressors that link to the high pressure gas storage and then to the 
freeze-out purifier inside the liquefier cold box before being reused.
The total amount of helium in the system is monitored in order to track 
losses~\cite{ziemann2016}.

%
%
\section{Radio-Frequency Powering System}

FREIA developed two prototype RF stations for testing ESS double-spoke cavities 
at 400\,kW, 352\,MHz, and corresponding RF distribution. R{\&}D work is ongoing
for the development of solid-state amplifier technology.

\subsection{RF Power Stations}

\begin{table}[t]
\centering
\caption[RF Station Parameters]{Main parameters of the tetrode RF stations.}
\label{tb:rfsource}
\begin{tabular}{l l}
\multicolumn{2}{l}{ } \\
Frequency                       & 352.21\,MHz \\
Bandwidth at 3 dB               & $>$250\,kHz \\
Repetition rate                 & 14\,Hz \\
RF Pulse length                 & 3.5\,ms \\
Maximum peak output power       & 400\,kW \\
RF Output connection            & 6-1/8\,inch, coaxial, 50\,$\Omega$ \\
Cooling                         & water (anode) and air (filament) \\
\end{tabular}
\end{table}

\begin{figure}[b]
\ifexpandfigures
   \centering
   \includegraphics[clip,keepaspectratio,width=\textwidth]{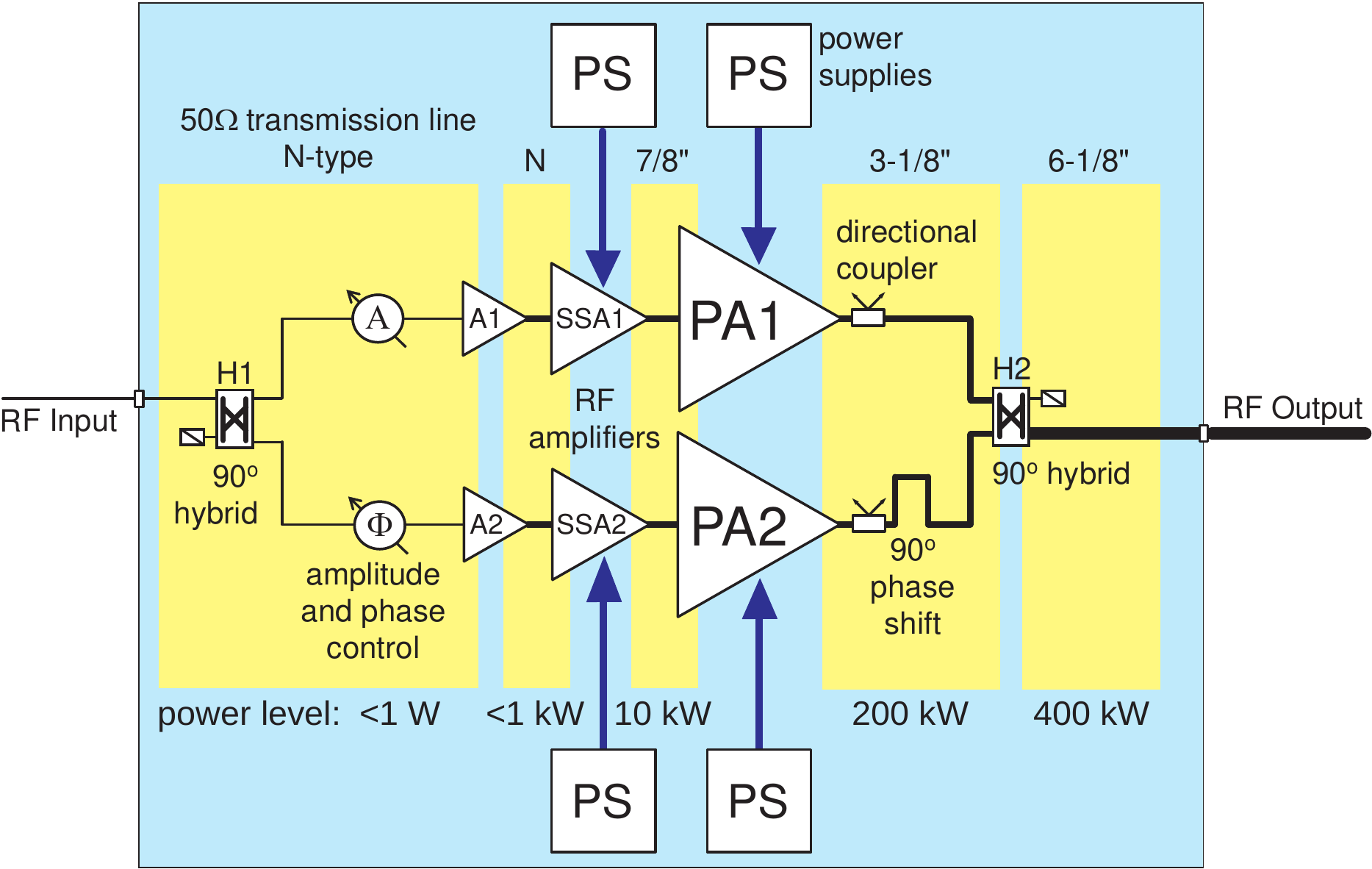}
\fi
  \caption[RF Amplifier]{\label{fig:rfsource}
  Schematic of the RF power station showing the RF signal input (left), split 
	in two paths leading via two pre-amplifier stages (A1, A2, SSA1, SSA2) to the
	tetrode amplifiers (PA1, PA2).
	Power levels, phase differences, coaxial connection types and other ancillary
	components are indicated.
	}
\end{figure}

\begin{figure}[b]
\ifexpandfigures
   \centering
   \includegraphics[clip,keepaspectratio,width=\textwidth]{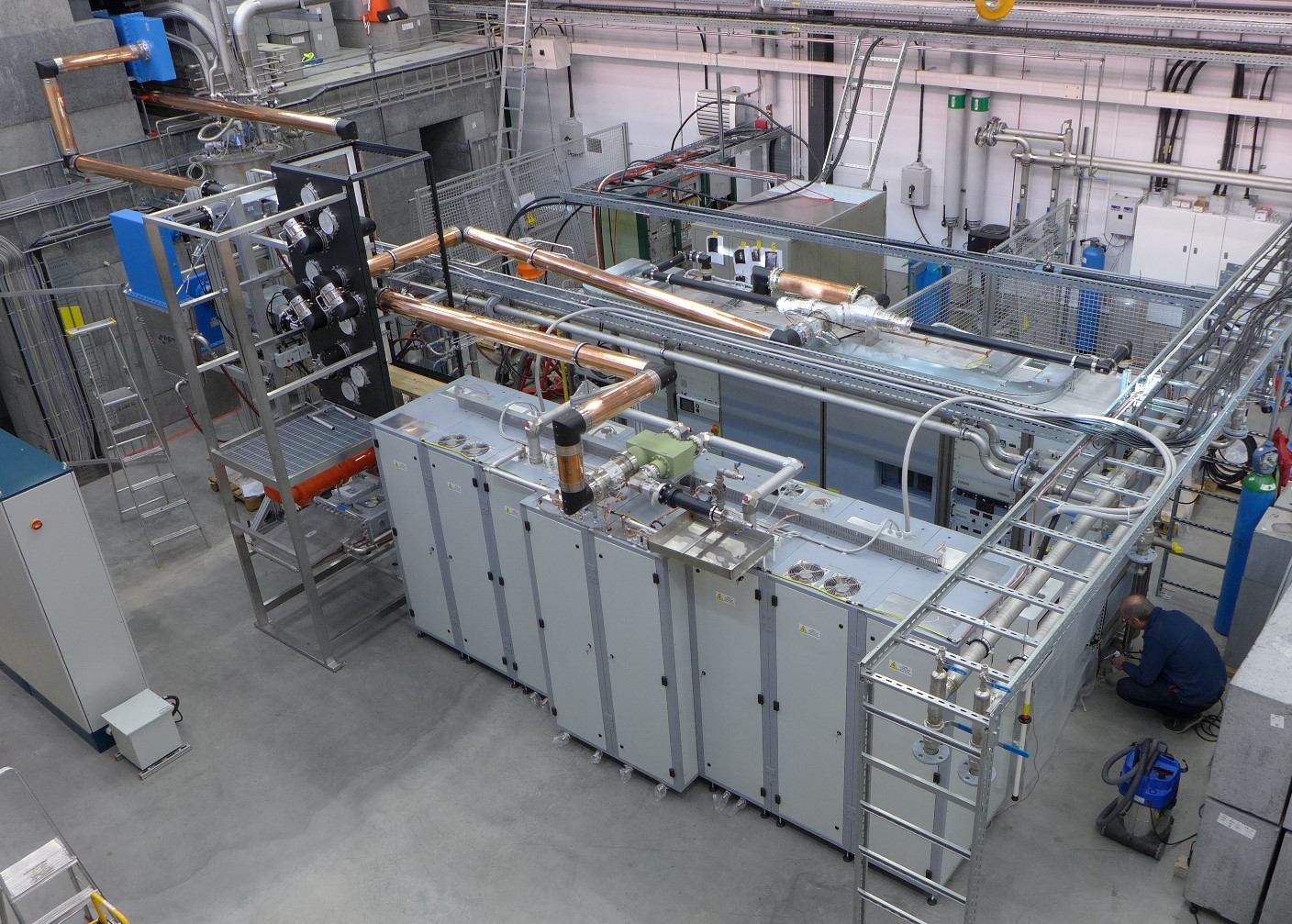}
\fi
  \caption[RF Stations]{\label{fig:rfstations}
  Photo with the two tetrode based RF stations (gray 8-door racks) in the 
	centre and the bunker on the left. Coaxial RF transmission lines (copper 
	coloured) come out from the top of the RF stations and go to the switch panel
	(black) from where they are routed to the bunker. One coaxial line enters
	directly and the other is connected to a (blue) waveguide transmission line.
	}
\end{figure}

The RF power stations were developed as prototype for the ESS. The design was
made in 2012 and combines two tetrode amplifiers for an output power of 400\,kW
pulsed at 14\,Hz~\cite{freia:report1204}.
The main parameters are listed in table~\ref{tb:rfsource} and the schematic of 
the tetrode based RF power station design is shown in figure~\ref{fig:rfsource}.
The low-power RF input from a signal generator enters from the left. A hybrid splitter (H1) divides this low-power RF signal into two equal-amplitude signals
with a 90$^\circ$ phase difference which each pass through individual
amplification channels. One of the signal channels incorporates amplitude 
regulation and the other phase regulation to balance the signal amplitude and
phase at the output of the amplifier chains. 
Then the signals pass through solid-state amplifiers A1 and A2 followed by
amplifiers SSA1 and SSA2~\cite{yogi2012}. The combined
amplification of these solid-state amplifiers is in the order of 70\,dB reaching
a signal level around 10\,kW. This signal then enters the final amplification
stage, the tetrode amplifiers PA1 and PA2. These have a minimum gain of 13\,dB, 
a conservative estimate as it is expected that the amplification gain of the
tetrodes deteriorates with ageing. After the amplifiers, the 90$^\circ$ phase
difference due to the input splitter H1 is compensated in order to match the
phases of both signals in the hybrid combiner H2 producing the output signal.
The hybrid combiner has a 0.15\,dB loss so the total amplification of the system
is about 80\,dB and a RF output signal in the order of 400\,kW is achieved.

Both signal channels contain several directional couplers for monitoring the
signal phase and amplitude.
As the power levels in the amplifier chain gradually increase, the size of the
transmission line increase from N-type coaxial cables to 7/8\,inch coaxial line
at the output of the solid-state amplifiers, to 3-1/8\,inch at the output
of the tetrode tube amplifiers, and finally 6-1/8\,inch after the hybrid 
combiner. The tetrode amplifiers are protected from reflected power by a
circulator at the RF output which is not part of the RF power station itself.

Each tetrode tube has four power supplies: one for the filament, screen grid,
control grid, and anode respectively. The anode power supply is a high voltage
power supply providing a voltage up to 18\,kV based on a capacitor bank and
charging unit~\cite{freia:report1204}. The high voltage anode power supply is
protected against short circuit by a crowbar and in one of the RF stations also
by a series switch circuit. The control grid power supply is pulsed between two
values: one for nominal operation of the tube and one for blocking the anode
current in-between RF pulses which increases the energy efficiency.

An initial survey revealed that the most suitable tetrode tube available was
the TH595 \cite{freia:report1201}. It can operate at power levels up to 200\,kW 
in pulsed mode with a gain of 15\,dB and an efficiency of 65{\%}. This tube has
a water-cooled anode and air-cooled cavity and tetrode connector, and can 
sustain an anode dissipation up to 40\,kW continuous operation. It is used with
a 352\,MHz output cavity named TH18595A. Recently an improved tetrode tube 
version TH595A has become available with a new screen-grid design allowing a
higher heat load and thereby enabling a possible higher duty cycle under ESS
specification in case of a future power upgrade of the ESS.
Thermal cycling effects due to switch-on and switch-off decrease the life time
of the tube. Therefore, a ``dark heating'' operation mode is used during long
periods of standby time in which the filament voltage is decreased 
substantially while avoiding a full thermal cycling of the filament.

Two tetrode RF stations based on our design were ordered from industry.
The initial design of the first RF power station was adapted for the second RF
power station adding an option for continuous operation and pulsed operation at
28\,Hz repetition rate. A separate solid-state pre-amplifier was included for 
the continuous operation in parallel to the pulsed-mode pre-amplifiers.
The two RF power stations, shown as the two 8-door racks in the center of
figure~\ref{fig:rfstations} were installed at FREIA in 2015 and 2016 
respectively and have been in operation since then 
\cite{freia:report1507,freia:report1701}. 
Based on this development, ESS chose a similar layout for their spoke RF power
stations.

\subsection{RF Power Distribution}

The peak and average power levels for the 352\,MHz RF power stations are such 
that coaxial transmission lines can be used, visible as the copper tubes on top
of the RF stations in figure~\ref{fig:rfstations}. These are 6-1/8\,inch 
coaxial lines as compared to more bulkier 23\,inch wide (half-height WR2300)
waveguide transmission lines chosen for example for the ESS accelerator
tunnel~\cite{freia:report1302}. A coaxial circulator for 400\,kW peak power
was prototyped and is now also used at ESS.

\begin{figure}[b]
\ifexpandfigures
   \centering
   \includegraphics[clip,keepaspectratio,width=\textwidth]{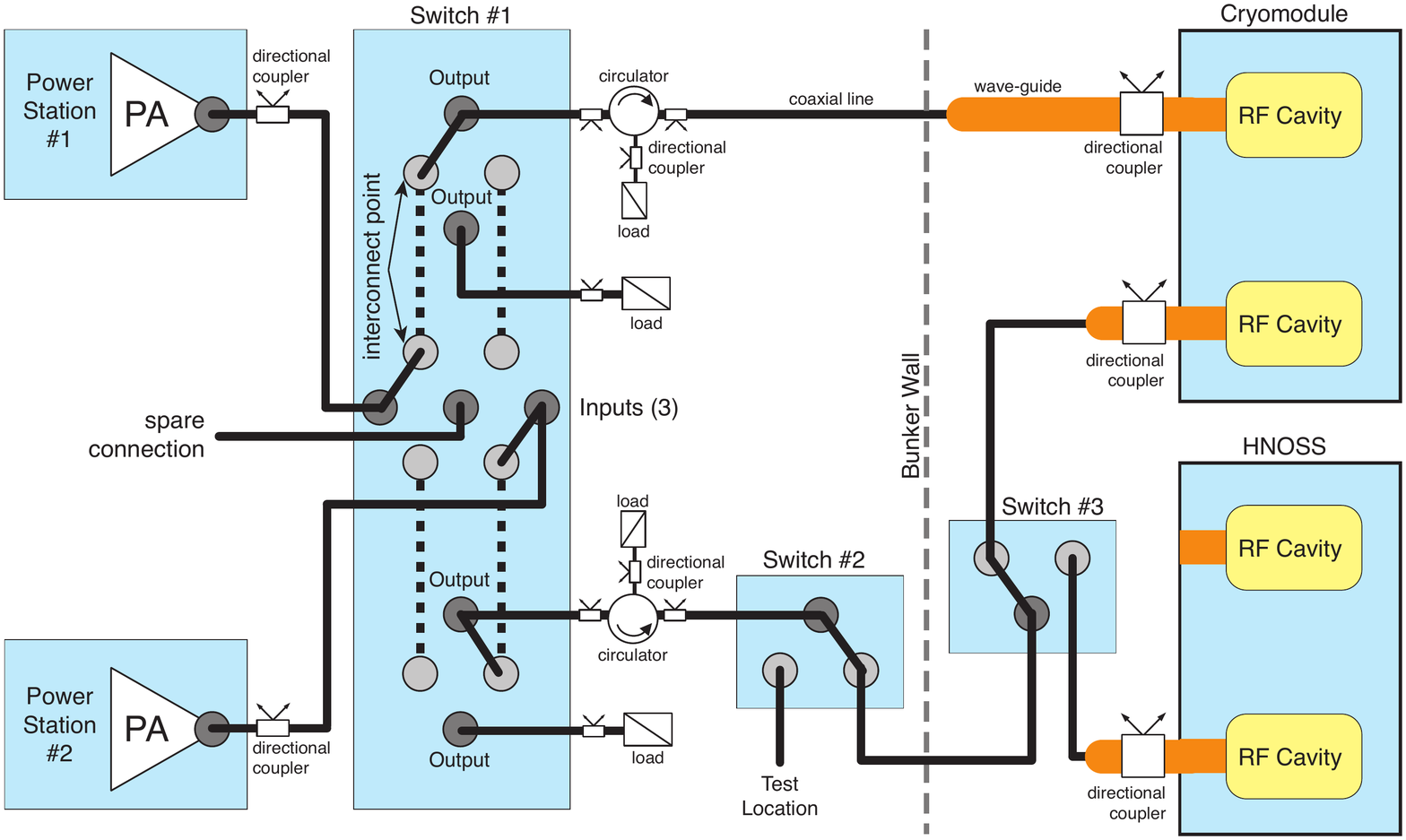}
\fi
  \caption[RF Distribution]{\label{fig:rfdist}
	Layout of the RF distribution from the top of the RF power stations (also
	shown in figure~\ref{fig:rfstations}) through coaxial lines to the switch
	panel and further on to the bunker either via coaxial lines (black) or
	waveguides (orange).
	}
\end{figure}

Having several RF power stations, connected to multiple locations, demands a
flexible solution for the RF power distribution. This was solved by connecting
the 6-1/8\,inch coaxial transmission lines from the RF power stations to a 
manual switch panel with 3 inputs, 4 outputs. The switch panel has 8 internal
connectors which allows any possible input-to-output combination using
fixed-length U-shaped coaxial-line connectors together with 4 permanent 
interconnections. As shown in figure~\ref{fig:rfdist}, the four outputs of the
main switch panel (marked switch \#1) can be connected to either a load, cavity,
or other test location. A circulator is installed on each connection to a test
location. The outputs to a load are used when testing and tuning the RF power
stations. Two smaller switch panels (marked switch \#2 and \#3) increase the
switching options. The switch panels are equipped with interlocks so the control
system can monitor the switch positions.

The ESS double-spoke cavities are equipped with a WR2300 half-height waveguide
transmission line as input. Therefore coaxial-to-waveguide transition are
installed in the final few meters before the cavity. The RF distribution system
is equipped with directional couplers to monitor forward and reflected power
levels.

\subsection{Solid-state Amplifier Development}

\begin{figure}[b]
\ifexpandfigures
   \centering
   \includegraphics[clip,keepaspectratio,width=\textwidth]{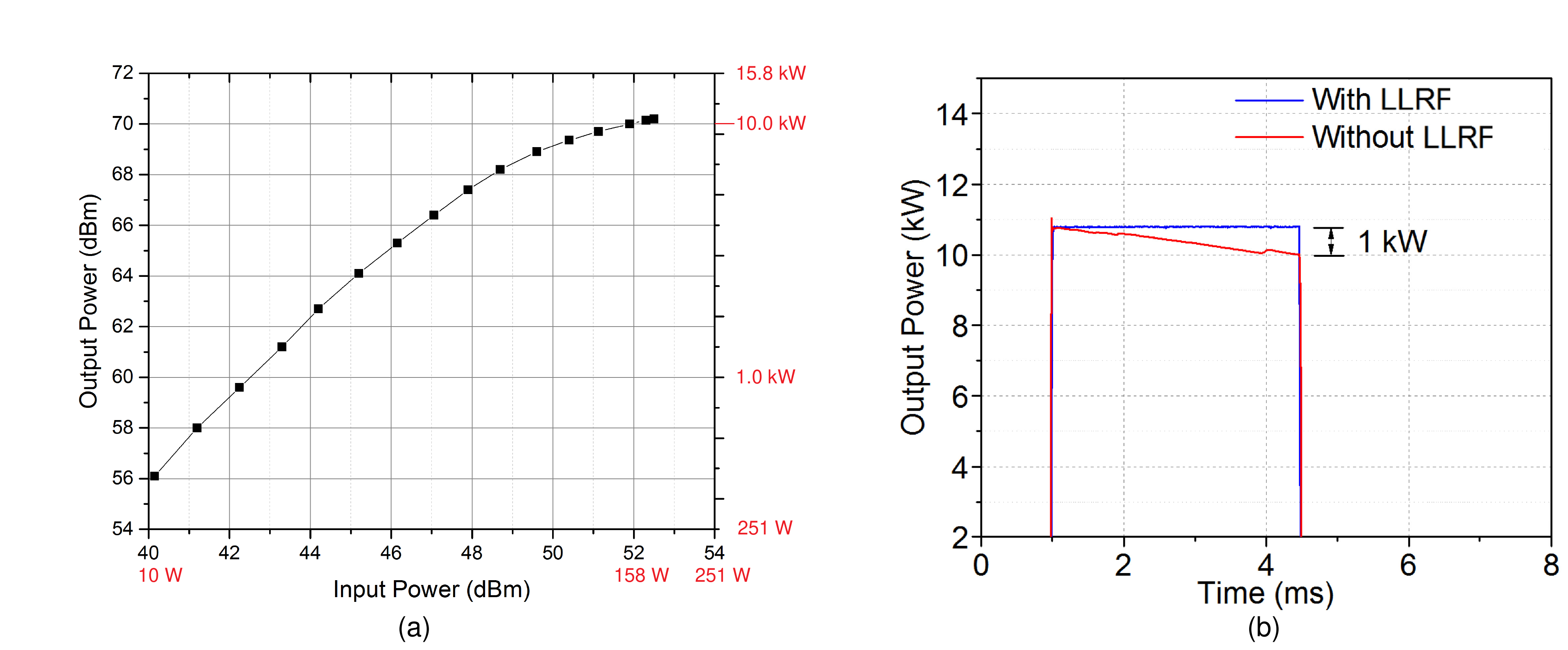}
\fi
  \caption[SSA]{\label{fig:ssa}
  Performance of the 8 transistor technology demonstrator.
	(a) Output power as function of the input power in logarithmic scale.
	Maximum output 10.5\,kW (70.2\,dBm) at 0.178\,kW (52.2\,dBm) input.
	(b) Variation of the uncompensated power amplitude during the pulse is in 
	the order of 1\,kW, while the compensated variation is only 20\,W at 10\,kW
	output power.
	}
\end{figure}

The possibilities of solid-state transistors for use in high-power RF 
amplifiers, as demonstrated at SOLEIL and other accelerators 
\cite{marchand2007,marchand2017}, encouraged the development of technology demonstrating parts for an ESS type RF power station to replace the tetrode
amplifiers described above. It was decided to concentrate on the two of the
technologically important parts: the transistor amplifier module and the RF 
power combiner. 

A single transistor version was chosen for the amplifier module, optimized for
overall efficiency and ease of manufacturing. The design effort resulted in a
single-ended amplifier at 352\,MHz realized in a planar printed circuit board
technology and avoiding using complex circuits, such as baluns, while oﬀering
the possibility to make all components machine mountable. An output power of 
1.3\,kW at 71{\%} efficiency and 19\,dB gain was achieved while operating with 
ESS parameters~\cite{dancila2014}.

An eight-transistor based technology demonstrator was constructed and operated 
successfully with feedback compensation for an optimized output pulse stability
\cite{dancila2017,hoangduc2019}. To reduce losses, the output of each 
transistors is tuned within 0.5\,dB gain and 5$^o$ phase. A total output power
of 10.5\,kW was achieved as shown in figure~\ref{fig:ssa}. 
Proportional and integral loop controllers, with a 6\,{\textmu}s processing
time, compensate the droop of the RF output power within the pulse caused by
the capacitor bank powering the transistors. The variation of the uncompensated
power amplitude is on the order of 1\,kW, while the compensated variation over
the pulse is only 20\,W, which is equivalent to 0.2{\%} \cite{hoangduc2019}. 
No instabilities appeared during 4\,h test runs while the temperature 
stabilized at the transistor and combiner~\cite{jobs2018}. For a 
multi-hundred\,kW RF station, the RF output of several 10\,kW amplifier stages
can be combined by a 12-ports 100\,kW re-entrant cavity type combiner which can
feed directly into a waveguide connecting the RF station to the accelerator 
cavity~\cite{goryashko2014,goryashko2018}.

%
%
\section{Test Infrastructure}

A key part of the test infrastructure is the control system. This includes
slow controls, monitoring of the infrastructure and equipment, and the fast
controls of RF power regulation and measurement, sometimes also called the
Low Level RF system (LLRF). In addition there are interlock systems for the
protection of the superconducting cavities and magnets to be tested, as well as
for the personnel safety.

\subsection{Slow Control and Interlocks}

The slow control system at FREIA is based on EPICS (Experimental Physics and 
Industrial Control System) \cite{epics} and LabVIEW \cite{labview}. The EPICS
software is widely used at research institutes for the control of accelerators.
EPICS is a distributed system consisting of a number of so-called Input/Output
Controllers usually equipped with hardware interfaces to the controlled process,
computers running services like archiving, logging, alarm, and computers used as
operator consoles. All these units are interconnected using an ethernet network.

The EPICS based controls are used at the top level to provide the operator with
a common interface to connect together different systems which might have their
individual proprietary control system. This includes systems based on commercial
industrial programmable logic controllers (PLC) or home-build laboratory systems
based e.g. on Raspberry Pi hardware~\cite{ziemann2016}.

The LabVIEW based controls are typically used for more complex experiments and
control programs. The LabVIEW controls are used for the conditioning and testing
of superconducting cavities and cryomodules. These methods and procedures are
often modified for specific experiments to adjust to changing and evolving
requirements from the end-user.

\subsection{RF Controls}

At FREIA, the LLRF system has been developed based on National Instruments 
PXI-type hardware including so-called field-programmable gate arrays (FPGA) for
fast data processing. To operate the cavity, either one locks the resonance
frequency of the cavity, so the RF power station can be locked to the same
frequency, or the RF power station follows the changing resonance frequency of
the cavity~\cite{powers2007,graf1991}.
The first operation mode is achieved with a generator driven loop and the 
second with a phase-locked loop or a self-excited loop.

In the signal generator-driven loop, the cavity resonance frequency is locked
and a signal generator produces an RF signal with a fixed frequency, which is 
then amplified by the RF power station and send to the cavity. A signal
proportional to the accelerating field in the cavity is picked up by an antenna
in the cavity and then used by the control system to tune cavity phase and
field amplitude.

\begin{figure}[b]
\ifexpandfigures
   \centering
   \includegraphics[clip,keepaspectratio,width=\textwidth]{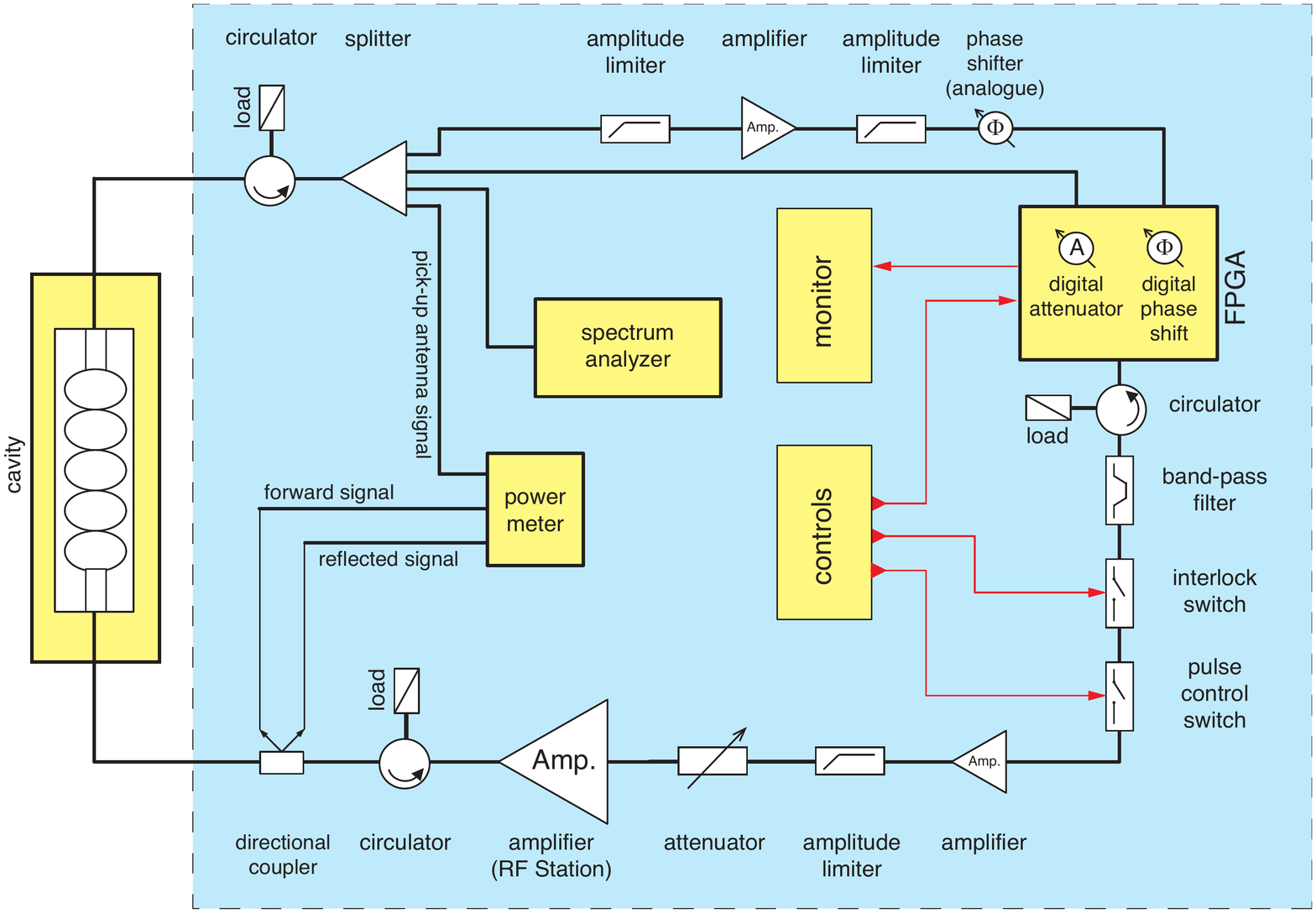}
\fi
  \caption[SEL]{\label{fig:sel}
  Diagram of the self-excited loop.
	The signal from the cavity antenna (top left) passes through signal
	preparation components to the analogue phase shifter and digital phase 
	shifter (top right) before passing through additional filters and amplifiers
	back to the cavity.
	The performance of the loop is mainly monitored in a time domain window and
	a frequency domain window in a computer connected to the FPGA unit, with in
	addition a spectrum analyzer and power meter.
	}
\end{figure}

In the self-excited loop~\cite{delayen1978,delayen2001}, a free-running RF
amplifier drives a high-gain positive feedback loop that is by nature unstable
and operates at a limit cycle defined by the cavity that functions as a narrow 
bandpass filter. The loop follows changes in the cavity's resonance frequency 
and therefore there is no need for an external frequency source and frequency
tracking feedback.
The self-excited loop developed at FREIA has a maximum gain of up to 100\,dB and
is operated with either an analogue trombone-type phase shifter or a digital 
phase shifter. A diagram of the self-excited loop is shown in 
figure~\ref{fig:sel}.
Before entering the phase shifter, the pick-up signal from the cavity passes
through an amplitude limiter and amplifier to ensure an adequate signal 
amplitude level. After the phase shifter, the signal passes through a bandpass
filter selecting the main loop frequency after which the signal is amplified 
and sent to the RF power station and back into the cavity. If we do not want to
operate the loop in continuous wave, the signal pulse length in the loop can be
controlled by a switch. Circulators are included throughout the loop to protect
amplifiers from reflected signals. An interlock switch is available to turn 
off the loop.

A quench detection algorithm is integrated into the digital RF control system
that regulates the signal generator-driven and self-excited loop. This is
especially of importance for the self-excited loop as during a cavity quench,
the cavity's resonance frequency changes and the self-excited loop will follow 
the drift in frequency and keep powering the cavity. The algorithm constantly
monitors the loaded-Q value of the cavity. In pulsed operation of the loop, the
FPGA, that processes the data from the loop, calculates the slope of the RF
field decay at the end of each pulse. If the exponential time constant, which
is equal to the inverse of the decay rate, is smaller than a pre-calculated or
measured value, it is assumed that a quench is starting to develop in the
cavity. In this case, the control system will engage an interlock to interrupt
the RF signal to the cavity.

%
%
\section{Recent Results}

Several cavities and cryomodules were tested at the FREIA Laboratory.
Cavities tested in Hnoss were fitted with a jacket serving as liquid helium
vessel. The first two cavities, a single-spoke and a double-spoke, were tested
with an input power in the order of 100~W, which we call low-power tests.
The next two cavities, a double-spoke and an elliptical, were tested at an input
power above 100~kW, which we call high-power tests.
The Gersemi vertical cryostat has been used for testing a crab cavity without a
liquid helium vessel.

\subsection{Single-spoke Cavity Low-power Test}

The first cavity test in Hnoss was a low-power test of a single-spoke cavity
borrowed from IJCLab. It was used to commission the cooling and test methods
including the used hardware set-up such as controls and self-excited loop.
The first step was to understand and develop appropriate calibration procedures 
required as an important step for any accurate measurement~\cite{freia:report1510}.
For example, as the attenuation of the copper cables for RF measurements depends
on their temperature, all calibrations had to be performed after cool-down of
the cavity to its 2\,K operating temperature~\cite{freia:report1602}. The cable
calibration was performed using a vector network analyzer and RF power meters.
Then, using the self-excited loop, it was possible to proceed making accurate
measurements of the cavity's resonance frequency, Q-factor, and accelerating
gradient.
Based on the test results, verified with previous test results from IJCLab, 
we developed an automatic test interface to improve the test efficiency and 
could, for example, shorten the Q measurement from two days to twenty minutes.

\subsection{Double-spoke Cavity Low-power Test}

\begin{figure}[b]
\ifexpandfigures
   \centering
   \includegraphics[clip,keepaspectratio,width=\textwidth]{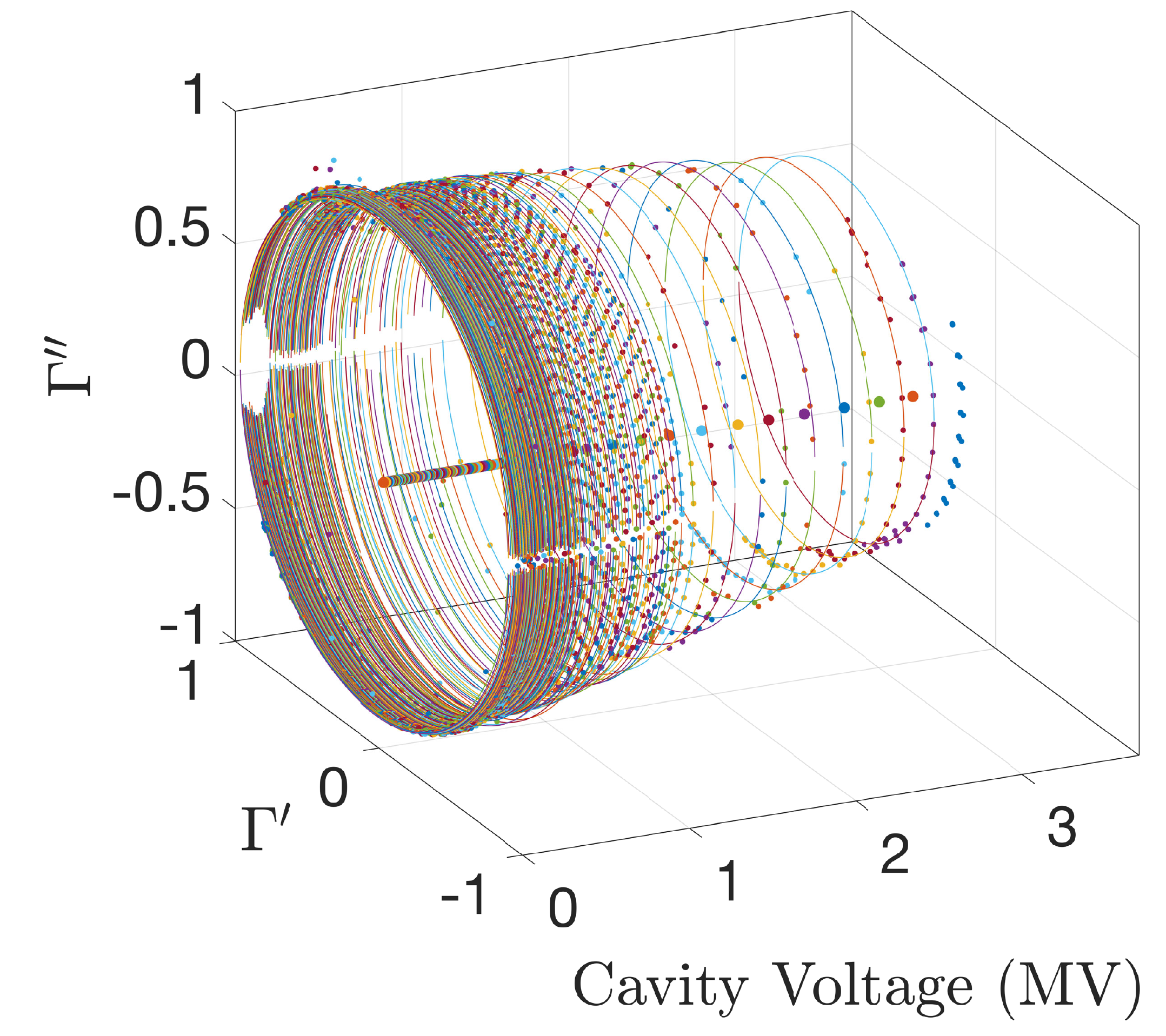}
\fi
  \caption[Q surface plot]{\label{fig:qsurface}
	Q surface plot with the measured real and imaginary parts of the reflection
	coefficient versus cavity voltage~\cite{bhattacharyya2018}.
	The data points of the same color are obtained for fixed forward power to the
	cavity but different phase shifts through the self-excited loop.
	The circles are the guide to the eye for data points grouped into thin slices
	(Q-circles) having approximately the same cavity voltage.
  }
\end{figure}

The second test in Hnoss was a low-power test of a 352\,MHz double-spoke cavity,
also from IJCLab, which was used to test the self-excited 
loop~\cite{freia:report1601}.
A high-precision Q-surface method was developed to accurately measure the Q$_0$
quality factor of the cavity as function of the cavity voltage or accelerating
gradient \cite{goryashko2016}. Typical uncertainty of the Q$_0$ factor found in
this way is in the order of 10 to 15{\%}. The standard Q-factor measurement
suffers from a deficiency originating from a single data point measurement of 
the reflection coefficient. With a Q-surface method, we are able to improve the
accuracy by an order of magnitude. In order to obtain the Q-surface, for fixed
forward power to the cavity we change step-by-step the phase shift across the
cavity by tuning the digital phase shifter installed in the self-excited loop.
This procedure is performed for each power level of interest, yielding the 
complex reflection coefficient of the cavity as a function of the cavity voltage
and phase shift as shown in figure~\ref{fig:qsurface}. The measurement points 
are grouped into slices around the same cavity voltage value, with each slice 
of measurement points forming a circle in the two dimensional space of the 
complex reflection coefficient. An accurate calculation of Q$_0$ is then 
obtained from the circle radius by means of a least-square minimization.

\subsection{Double-spoke Cavity High-power Test}

\begin{figure}[b]
\ifexpandfigures
   \centering
   \includegraphics[clip,keepaspectratio,width=\textwidth]{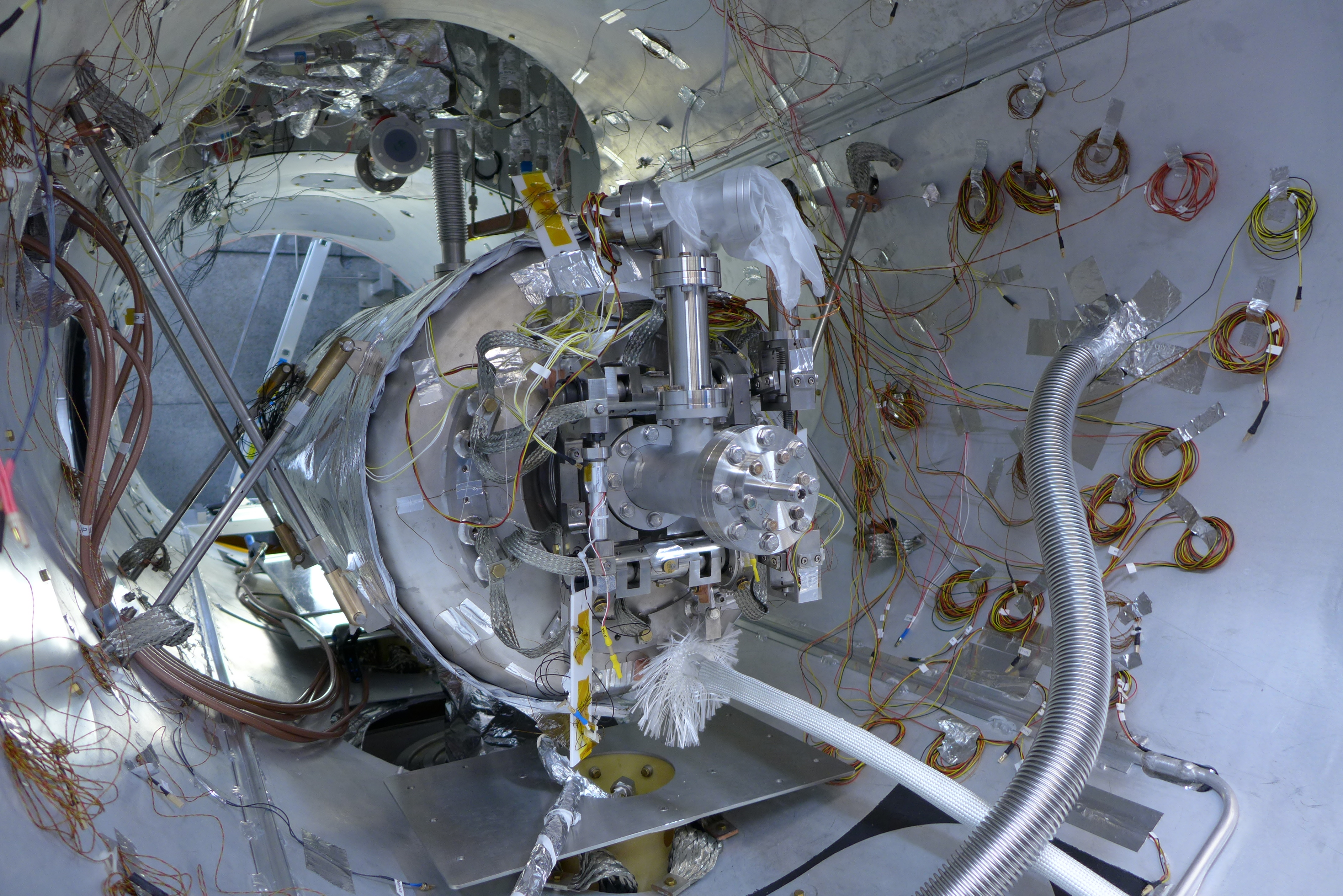}
\fi
  \caption[Cavity in Hnoss]{\label{fig:cavity-install}
  On-going installation of a double-spoke cavity in Hnoss, hanging from the
	vacuum vessel wall by support rods. The tuner is mounted at the near end and
	surrounds the beam pipe. Many signal cables are still taped to the inner wall
	of the thermal shield.
	}
\end{figure}

The third test in Hnoss was the first one using the high-power RF stations.
The double-spoke cavity, equal in design to the previous cavity tested at 
low-power, was equipped with liquid helium jacket, high-power RF input coupler,
and frequency tuning system. As such it was a completely equipped prototype for
the ESS accelerator~\cite{freia:report1710}. 
Figure~\ref{fig:cavity-install} shows a photo of its installation in Hnoss.

The RF input-coupler was conditioned at room temperature and then twice at
cryogenic temperature (cavity at 2\,K). The first cold conditioning was done with
an LLRF system borrowed from IJCLab. The second cold conditioning was done 
with the FREIA LLRF system. The RF station power level is controlled by the
automatic conditioning software, while all essential safety interlocks are
implemented in hardware.

RF power conditioning is done in standing wave mode with pulse length increasing
from 20 to 2860\,{\textmu}s. During each step of selected pulse length, the
power ramps up slowly in steps depending on various parameters set in the 
software until the maximum defined RF power is reached.
Two software vacuum thresholds are used: as long as the vacuum pressure measured
at the power coupler is below the first threshold, the RF power is step-wise
increased. When the vacuum pressure rises above the first software threshold,
the RF output power is kept stable until the vacuum pressure recovers. 
Otherwise, if the vacuum pressure increases above the second threshold, the RF
power is decreased until the vacuum pressure decreases below the first 
threshold. The RF frequency used during the power coupler conditioning is just
outside the resonance frequency of the cavity to avoid RF fields building up
inside the cavity. 

Next the cavity is conditioned using a pulsed self-excited loop mode.
First, at very low power, a digital phase shifter is used to vary the loop 
delay, and thereby the frequency, around the cavity resonance frequency. This
creates a controlled sweep of the field distribution back and forth along the
coupler walls.
In the second phase the RF power is ramped up while keeping a fixed pulse 
length. After about 30~hours of conditioning and passing through three 
major multipacting regions, the cavity package reached a stable 9\,MV/m flattop
accelerating gradient. This is slightly above the nominal gradient for this 
cavity design intended to be used for the ESS accelerator~\cite{li2019}.

\subsection{Elliptical Cavity High-power Test}

The fourth test was a high-power test of a prototype 704\,MHz elliptical cavity
for the ESS accelerator. The cavity was equipped with high-power RF input 
coupler and frequency tuning system. It was tested to verify design and 
operation of the high-power coupler. RF pulses with different amplitude, 
duration, and repetition rate recreate a similar situation as the cavity would
experience in the accelerator. Overall, a similar test program was performed as
for the high-power test of the double-spoke cavity.

For this test we installed a 704\,MHz klystron amplifier powered by a 
high-voltage pulse modulator, or modulator in short.
We first commissioned the modulator followed by the klystron up to 1\,MW peak power, limited by the RF load maximum power dissipation.
The RF system and cavity conditioning was done with a LLRF control system
developed by Lund University for use at the ESS accelerator, and with our own
automatic conditioning system software.

The cavity was cooled down to 2\,K two times without any significant issues. 
The power coupler and the cavity were successfully conditioned at high power RF.
An accelerating gradient of 13\,MV/m was measured with an incident peak RF power 
of 290 kW. Some outgassing and X-ray radiation occurred at intermediate field
levels, but disappeared after a few hours of operation. The effect of the
Lorentz force detuning was visible at high field, with a variation of the field
amplitude on the flat top of the pulse~\cite{freia:report1807,li2018}.

Two different methods are used for dynamic heat load measurements.
The first method is by liquid helium evaporation, measured via a gas flowmeter
placed after the sub-atmospheric pumps. The second method is by pressure rise, 
and is also used to cross check the system performance.
The liquid helium volume of the cavity is closed off and the rise of helium
pressure is measured as function of time. A heat load calibration curve was 
prepared by repeatedly measuring the relative rise of the pressure level caused
by a known heat load from a resistive heater.

\subsection{Spoke Cryomodule High-power Test}

\begin{figure}[b]
\ifexpandfigures
   \centering
   \includegraphics[clip,keepaspectratio,width=\textwidth]{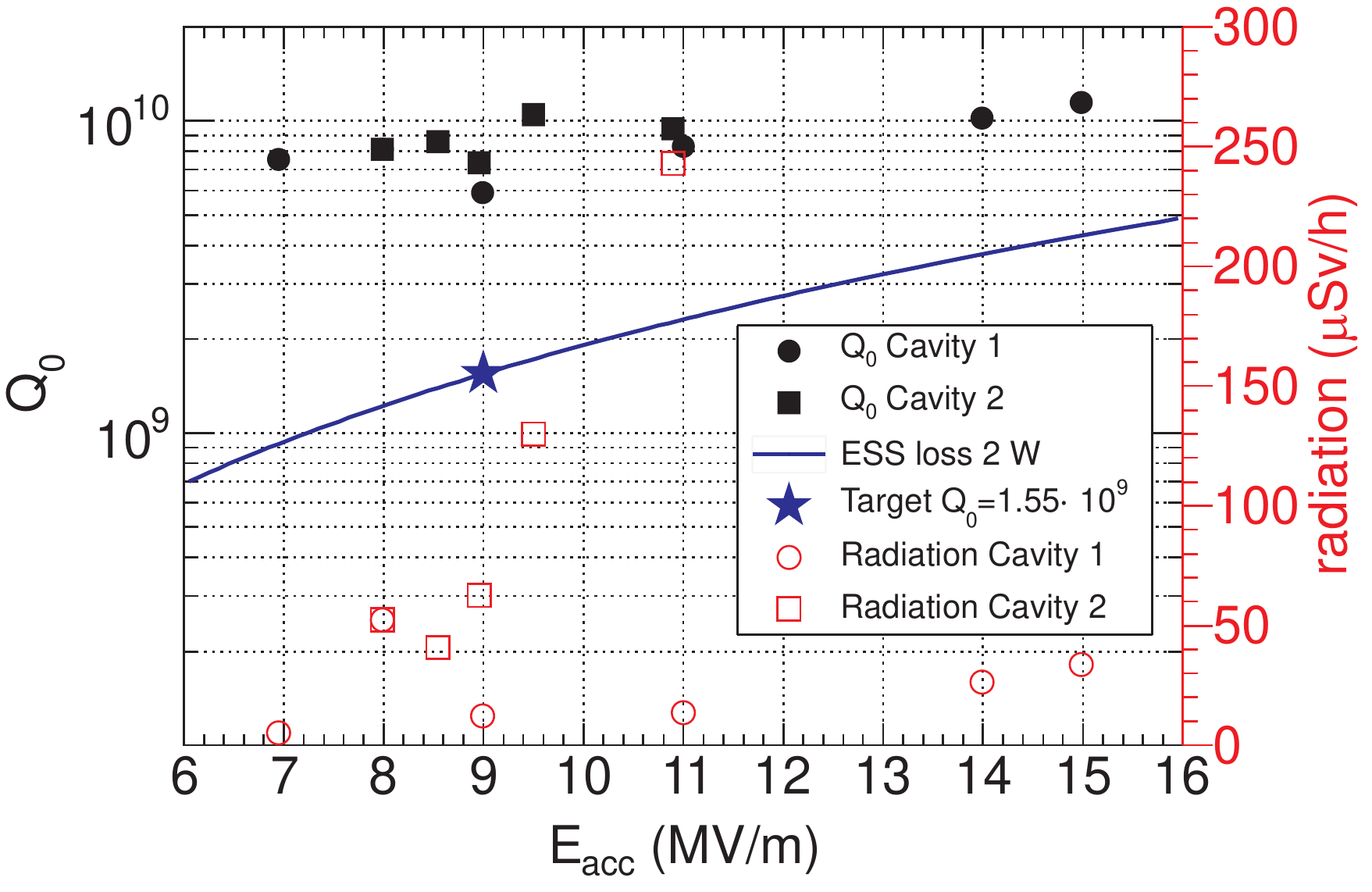}
\fi
  \caption[Prototype CM Q-factor]{\label{fig:dsrqfactor}
  Measured Q factor of the two cavities in the prototype cryomodule.
	Data is compared to the ESS target value (star), measurements in a vertical
	cryostat before mounting in the cryomodule (courtesy IJCLab),
	and X-ray radiation levels recorded when performing the measurements.
	}
\end{figure}

A prototype cryomodule for the ESS accelerator~\cite{garoby2017} equipped with
two double-spoke cavities was tested in 2019 with more optimized parameters 
based on the lessons learned from the earlier tests described above.
The test included studying the conditioning procedure, accelerating gradient,
dynamic heat load, cold tuner, piezo actuator, diagnostics, RF controls, and
safety interlocks.
A special cryogenic valve box was installed next to Hnoss, in the same bunker, 
and connected in parallel onto the cryogenics distribution system using the 
same cryogenic interconnection box as for Hnoss.

The conditioning procedure was divided into power coupler conditioning and 
cavity conditioning~\cite{miyazaki2020}. 
First, the power couplers were conditioned at room temperature before the 
cooling of the cryomodule was started. Second, the power couplers were
reconditioned after the cavities were cooled down. Finally, the cavities were
conditioned. Every time a thermal cycle occurred, for example for dedicated 
tests of the cryogenic system, these three steps were repeated to recover the
functionality of the cavities.

The cavity conditioning started in pulsed self-excited loop mode. A second
conditioning of the cavities was performed with the generator-driven loop. The
conditioning was carried out with 1 and 3.2\,ms pulse lengths while slowly
increasing the RF power and hence the accelerating gradient.
Figure~\ref{fig:dsrqfactor} shows the measured quality factor of both cavities
as function of the accelerating gradient. Electron activities in the cavity 
were associated with X-ray dose rate and given only for a relative comparison
of the level of radiation as function of the accelerating field. Both cavities
in the cryomodule achieved the ESS nominal operation parameters.

The cryomodule was thermally cycled several times. Noticeable outgassing
occured every time the cavities warmed up above 20\,K, with temperature and
pressure ($<$10$^{-8}$\,mbar) implying that nitrogen was the dominant vapour
element. A small thermal cycle of the cavities from 2 to 50\,K did not affect
the conditioning. After a full thermal cycle to room temperature for one month,
the re-conditioning of the couplers required only a few hours. However cavity
re-conditioning took almost the same time as the first conditioning after the
first cool down~\cite{freia:report1905,freia:report1908}.

\subsection{Crab Cavity Low-power Test}

\begin{figure}[b]
\ifexpandfigures
   \centering
   \includegraphics[clip,keepaspectratio,width=0.5\textwidth]{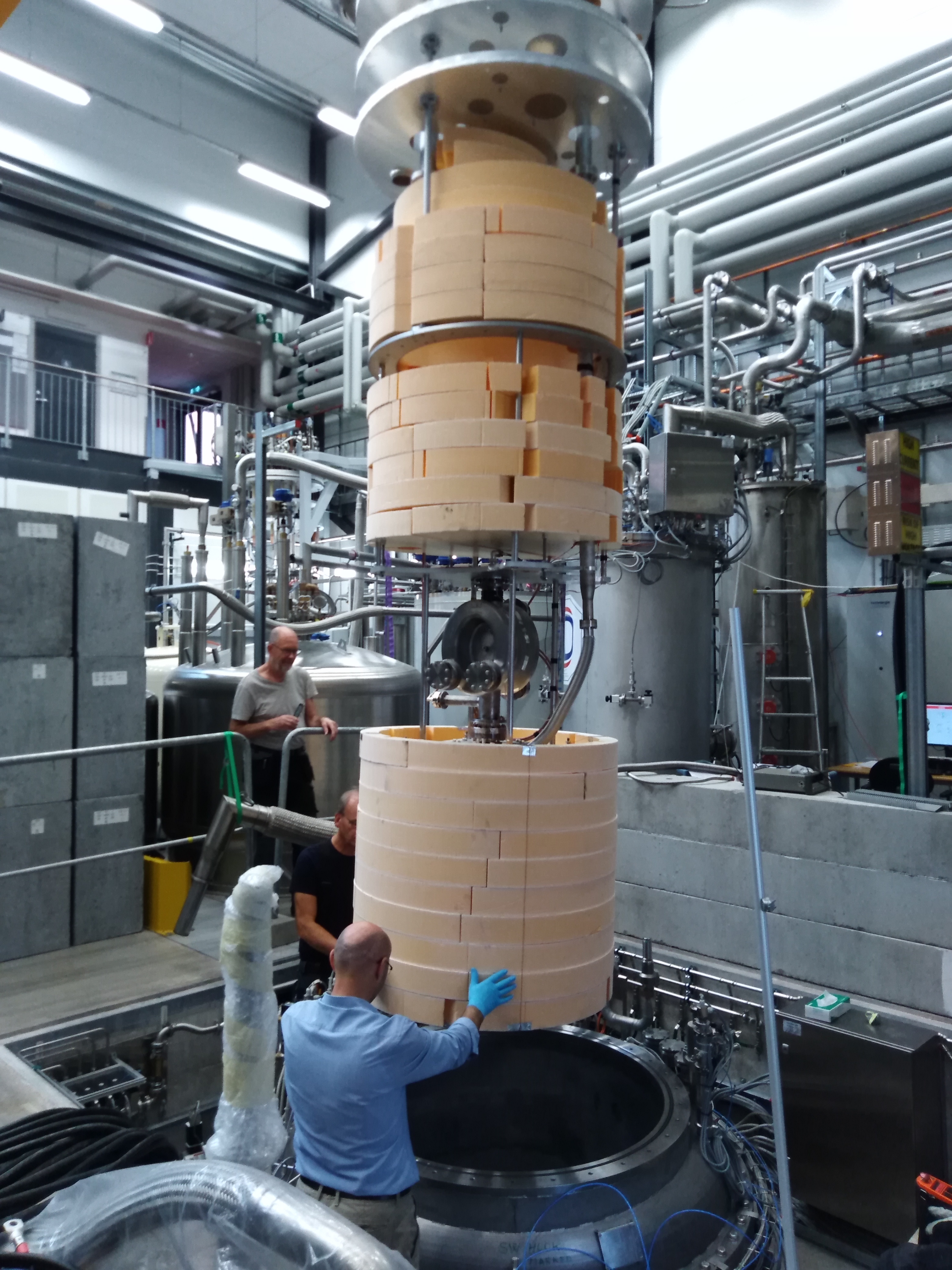}
\fi
  \caption[Crab Cavity]{\label{fig:dqwphoto}
  Lowering the insert with an HL-LHC crab cavity into Gersemi.
	Note the orange coloured foam to decrease the amount of liquid helium
	required to fill the large cryostat volume.
	}
\end{figure}

\begin{figure}[b]
\ifexpandfigures
   \centering
   \includegraphics[clip,keepaspectratio,width=\textwidth]{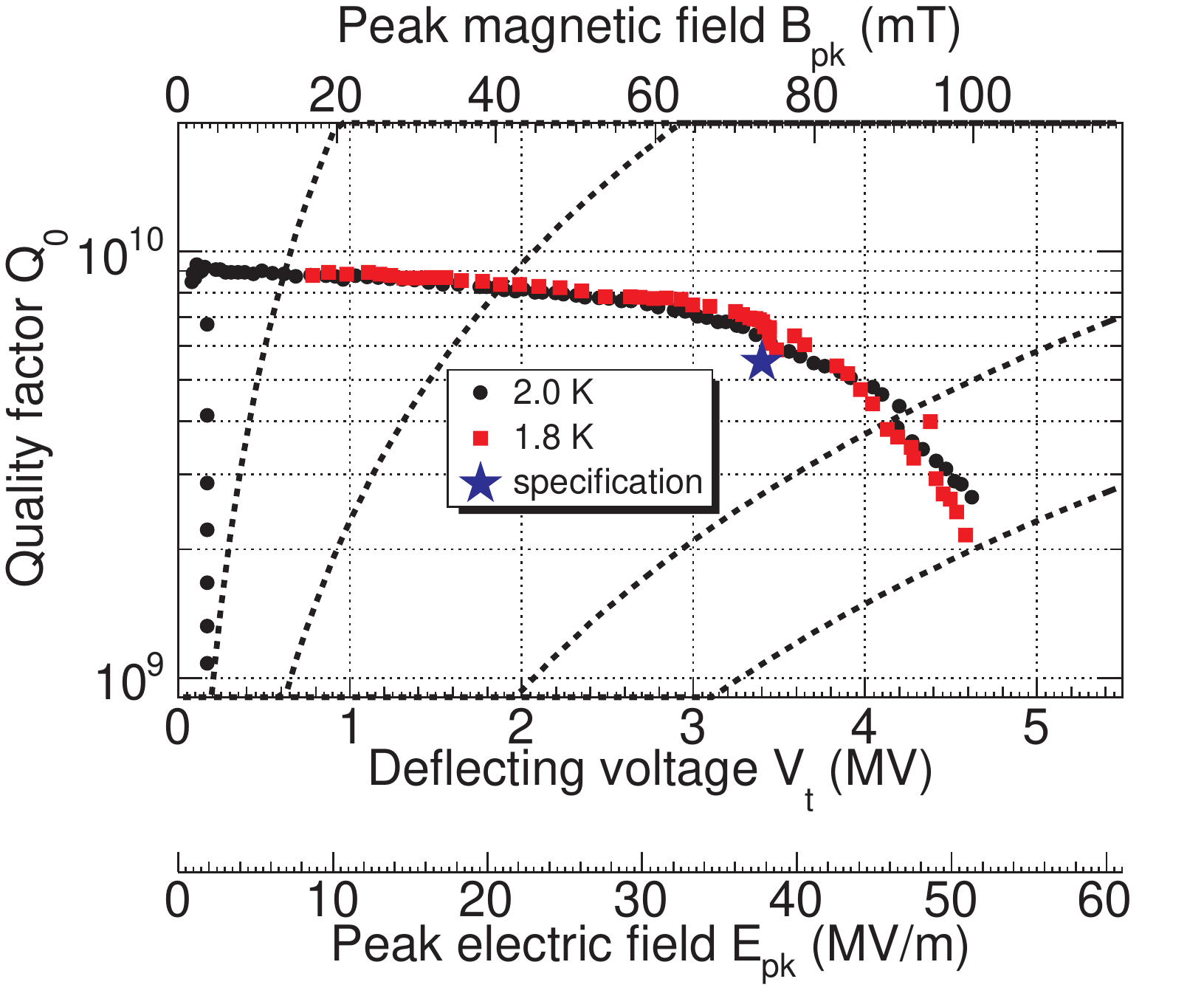}
\fi
  \caption[Q versus V]{\label{fig:dqwtest}
  Measured quality-factor Q versus transverse voltage V$_\mathrm{t}$ for the
	HL-LHC crab cavity prototype. The nominal specification value is indicated by 
	a star. Dotted lines indicate RF power levels in the cavity.
  }
\end{figure}

A 400\,MHz crab cavity prototype for the High Luminosity upgrade of the 
Large Hadron Collider (HL-LHC) was the first cavity tested in the Gersemi 
vertical cryostat. 
This cavity was a double quarter wave type for horizontal bunch crossing
in the ATLAS detector of the LHC, operated in continuous mode~\cite{verdu2018}.
The cavity was manufactured in the USA and previously tested at Jefferson
Laboratory and CERN. The test in Gersemi was performed as a first step to 
validate the ability of FREIA to test these cavities and serve as a back-up 
test stand for CERN and the HL-LHC project.
As the cavity came without its liquid helium jacket, it was tested in Gersemi
with operation mode and insert type 1 (figure~\ref{fig:gersemi-inserts}) and
supported by a solid mechanical frame.
Figure~\ref{fig:dqwphoto} shows the insertion of the cavity into Gersemi.

The cavity has been validated to operate above the nominal gradient of 3.4\,MV 
at 2\,K. Figure~\ref{fig:dqwtest} shows the measured cavity Q factor versus
the transverse RF voltage V$_\mathrm{t}$. The cavity was also measured at 1.8\,K 
to demonstrate the availability of Gersemi at both temperatures. The main 
challenge for the measurement was the high Lorentz force detuning of the cavity
which shifted the resonance frequency as RF power levels increased. The 
experiment was not performed with a self-excited loop but with a phase-locked
loop based on electronics lent by CERN. Cavity resonance was lost at 4.6\,MV
transverse voltage, both at 2\,K and 1.8\,K, which we interpret as the quench
limit~\cite{miyazaki2020a}.

%
%
\section{Current and Future Actitivies}

The different development projects and subsequent testing of superconducting
cavities has prepared the FREIA Laboratory for a wide range of accelerator
development activities. Some of them are briefly explained below.

\subsection{ESS Spoke Cryomodules}

At the time of writing, FREIA is testing a series of 13 spoke cryomodules for
the ESS accelerator. Like the prototype spoke cryomodule, each cryomodule is
equipped with two double-spoke cavities. The first cryomodule arrived at FREIA 
in October 2020 and testing was completed in December 2020. To keep the ESS 
high-level planning, each of the next cryomodules is to be tested within a 
tight schedule.
Turn-over is expected to reach an average of six weeks per cryomodule.
It takes about one and a half week for reception tests and installation, 
one week for warm RF conditioning of the power couplers of both cavities, 
then one and a half week for cool down, cold RF conditioning and cold testing,
one and a half week for warm-up and removal, and a few more days to prepare
for shipment.

\subsection{LHC High Luminosity Upgrade}

Key elements of the HL-LHC project at CERN are superconducting magnets and
cavities. FREIA is ideally suited to contribute to the development of such
equipment. Work is ongoing in three areas: superconducting dipole magnets,
superconducting crab cavities, and cryostats for the superconducting power 
link.

The Gersemi vertical cryostat is available both for superconducting magnet and
cavity testing. The superconducting crab cavities are similar in design to the
previously tested 400\,MHz crab cavity. They are tested before mounting their
liquid helium jacket and the test can therefore be performed in a simple liquid
helium bath. The HL-LHC project is also interested in cold power testing of 
some 20 superconducting dipole magnets used for beam orbit correction. These
magnets combine two dipole magnets, horizontal and vertical field direction, in
a single device with each having a dipole field strength of up to 2.2\,T in
different configurations with a length of either 1.5 or 2.1\,m.

The HL-LHC includes a superconducting link which has to ensure a 
current-carrying capacity of up to 100\,kA over a distance of 
100\,m~\cite{ballarino2014}. This superconducting link connects several
superconducting magnets to their respective power supplies by means of a vapour
cooled superconducting cable.
FREIA, in collaboration with CERN and Swedish industry, contributes to the 
construction of the cryostats that integrate and ensure cryogenic cooling with
helium vapour of resistive splices between MgB$_2$ and BSCCO type 
superconducting cables.

\subsection{Superconducting Canted-cosine-theta Magnet}

\begin{figure}[b]
\ifexpandfigures
   \centering
   \includegraphics[clip,keepaspectratio,width=\textwidth]{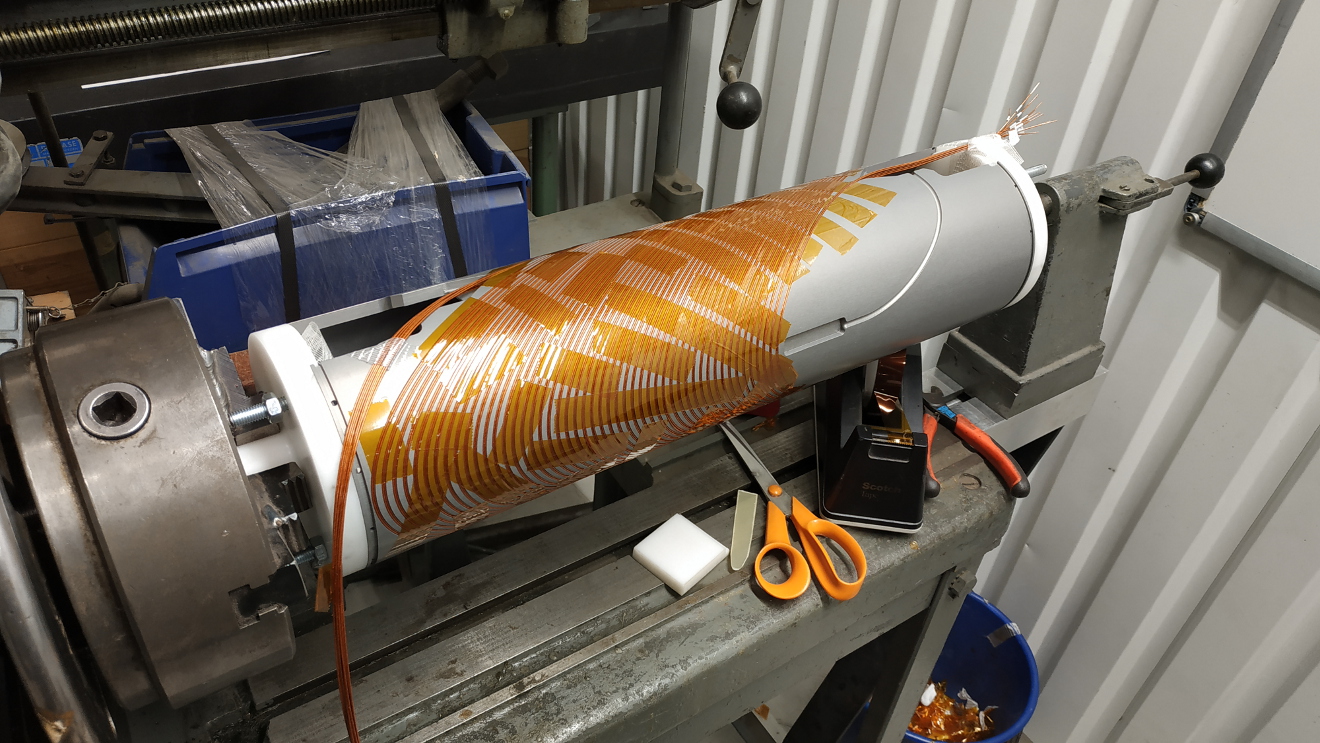}
\fi
  \caption[CCT]{\label{fig:cct50cm}
	Winding a coil for the 50\,cm long CCT model magnet.
	Photo courtesy Scanditronix Magnet AB.
  }
\end{figure}

In collaboration with Swedish industry, FREIA is developing a 
superconducting dipole magnet based on a canted-cosine-theta (CCT) layout,
sometimes referred to as double-helix or tilted coils. The CCT design is based 
on two concentric modulated helical coils having a winding shift of $\pi$ and
inverse currents~\cite{meyer1970}. 
Advances in the accuracy of 3D machining technology have now made it possible
to manufacture this kind of magnet layout by machining an accurate groove in
an aluminium former~\cite{kirby2018}.
Our CCT magnet is a dipole magnet build by the superposition of two oppositely
skewed solenoids with respect to the bore axis. A 50\,cm model dipole magnet,
shown in figure~\ref{fig:cct50cm}, was fabricated and is waiting to be tested 
in Gersemi.
Next we will start construction of a 1\,m long CCT type dipole magnet that
shall fulfil the specifications for a LHC type orbit corrector dipole magnet.

\subsection{Vacuum Breakdown and High-gradient Acceleration}

From the earlier participation in the Two-Beam Test Stand experiments at 
CTF3~\cite{TBTS} evolved an interest to probe the limits of electric field
gradients in accelerating cavities. We developed a spectrometer for the
measurement of dark and breakdown currents that is being operated at the CLIC
high-gradient test stands at CERN. By using tomographic image reconstruction it
became possible to obtain the vacuum breakdown location inside an accelerating 
structure~\cite{jacewicz2016}.

Recently, we have constructed a cryogenic, pulsed, high-voltage system 
integrated in a stand-alone cryocooler to investigate the fundamental mechanisms
of field emission and breakdown nucleation. In the system, high-field 
measurements can be  carried out at ambient and down to cryogenic temperatures.
The first results reveal a significant increase in the field holding capability
of the copper electrodes when cooled and conditioned at cryogenic temperatures,
between 30 and 90\,K~\cite{jacewicz2020}.

\subsection{Lasers and X-ray Sources}

Already before the establishment of the FREIA Laboratory, some of us were 
involved in the development of free-electron lasers (FEL). Noteworthy 
contributions are e.g. the Optical Replica Synthesizer for FLASH~\cite{ORS} 
and the Laser Heater for the European-XFEL~\cite{XFEL}. We are now working
actively on the design of soft X-ray FEL extension of the linac-injector at 
MAX~IV~\cite{mak2019,curbis2019} and participate to the Eu-supported
``XLS CompactLight'' design study for a compact hard X-ray FEL 
facility~\cite{dauria2019}. 
The gained knowledge is being combined to develop a design for a compact 
electron and X-ray source at FREIA. The source would contain a normal 
conducting RF gun and superconducting accelerating cavities to provide 3 to
10\,MeV electrons. At 3\,MeV it can be used for electron diffraction 
measurements, and at 10\,MeV to produce X-rays from Compton scattering.

\subsection{Microwave Amplifiers and Communication}

The work on the 352\,MHz solid-state amplifier development has been extended
to other frequencies and power levels. Designs and prototypes have been 
developed for power amplifier modules for cyclotrons in radioisotope production
that operate in the low VHF band at 27 and 100\,MHz~\cite{tong2019,tong2021}.

Research is ongoing on 60 GHz broadband antennas and transceiver chips for 
high-speed communication and data transfer in particle physics 
experiments. The upgrade of the Large Hadron Collider for high luminosity will
result in multiple times higher event rates, which demands high data rate 
readout systems in the order of 50 to 100\,Tbps. The feasibility of 60\,GHz
wireless links in harsh detector environments was studied by investigating
different antenna arrays at 28 and 60\,GHz and performance degradation of 
wireless communication chips after irradiation~\cite{aziz2019,aziz2019a}.

\subsection{Neutron Source}

A commercial deuteron-tritium type neutron source is being procured by 
colleagues at the Department of Physics and Astronomy. The source will provide 
14\,MeV neutrons with a yield exceeding $10^{10}$\,n/s. It will be used for
academic and educational purposes, as well as industrial testing of radiation
effects in electronics.

\subsection{ESS Upgrade}

The advent of the powerful ESS proton accelerator and the simultaneous interest
in neutrino physics gave rise to the idea to use the ESS accelerator to 
generate, concurrently with the generation of spallation neutrons, an intense
neutrino beam. The pulse repetition rate of the ESS accelerator will be doubled
and every second pulse will be diverted to a second target station to produce
a pion beam. The neutrinos from pion decay will be detected in a huge 
underground water Cherenkov detector ca. 500\,km from ESS to study neutrino
oscillations and leptonic CP violation. The muons from pion decay could also be
accelerated and brought into collision to produce a very large number of Higgs
bosons. ESSnuSB (ESS neutrinos Super Beam)~\cite{wildner2016} is a 4-years EU
supported Design Study of this research program with scientists from FREIA 
working on the design of an accumulator ring needed to compress the 3\,ms long
ESS linac pulses to 1.3\,{\textmu}s pulses.

\subsection{Other International Projects}

FREIA participates in the ARIES accelerator development project, a European wide
project with 42 participating universities and institutes~\cite{aries}. 
Our research on solid-state RF amplifiers is the FREIA part of the work package
related to improving the sustainability and efficient energy management of
accelerator infrastructures. The Hnoss and Gersemi cryostats and test
infrastructures are shared as a trans-national access facility, which
reinforced the case to construct the Gersemi cryostat for both superconducting
cavity and magnet testing. Several of the cavity tests described above were
performed in the framework of the ARIES trans-national access.
Within the new European I.FAST project, FREIA will lead a work package to 
develop a RF amplifier around 750\,MHz based on kW-level GaN transistors.

%
%
\section{Conclusions}

The FREIA Laboratory was initially conceived to develop part of the acceleration
system of the ESS. From there it has developed into a versatile infrastructure 
to support the larger accelerator community by ensuring the performance of 
high-end components for future accelerators. 
One of FREIA's focal points is the test of superconducting RF cavities and
magnets, which is facilitated by the availability of horizontal and vertical
cryostats, cooled by liquid helium, and high-power RF amplifiers. The latter is 
a key component for the development of the power generation, distribution and 
test of the spoke cavities for the ESS. In the past year, new fields of activity
opened up with tests of HL-LHC crab cavities and magnets, and the development of
canted-cosine-theta magnets. 
A second focal point is the development of solid-state power amplifiers and 
power combiners, which might well replace the RF vacuum-tube amplifiers that 
still power many accelerators today.
These main fields of activity are complemented by other projects, such as the
study of vacuum breakdown and the development of compact sources of radiation,
which may expand in the future. 
Parts of the FREIA infrastructure are internationally accessible as  
trans-national access facilities, which is expected to widen FREIA's scope 
further.

%
%
\ifjournallayout
  \acknowledgments
\else
  \section{Acknowledgements}
\fi

We like to thank our colleagues at the FREIA Laboratory and Uppsala University
for their support during this project. 
Our work would have been impossible to achieve without the support and advice
from our colleagues at
ACS Accelerators and Cryogenics Systems, 
CEA Saclay,
CERN, 
the European Spallation Source ESS,
IJCLab (former IPN Orsay), and
Lund University.

This project has received funding from 
the Knut and Alice Wallenberg Foundation,
the Vice-Chancellor and 
the Faculty of Science and Technology of Uppsala University,
and
the European Union’s Horizon 2020 Research and Innovation programme under 
Grant Agreement No~730871, 
No.~777419, 
and No.~777431. 

%
%

%

\ifjournallayout
  \bibliographystyle{JHEP} 
\else
  \bibliographystyle{unsrt}
\fi
\bibliography{freia-paper}

\newpage

%
%
\end{document}